\documentclass[5p]{elsarticle}

\pdfoutput=1
\journal{Journal of Process Control}
\makeatletter
\def\ps@pprintTitle{%
   \let\@oddhead\@empty
   \let\@evenhead\@empty
   \let\@oddfoot\@empty
   \let\@evenfoot\@oddfoot
}
\makeatother

\usepackage{natbib}
\usepackage{geometry}
\usepackage{fleqn}

\usepackage{amsmath, amssymb}
\graphicspath{{./Figures/}, {./Figures/Extended/}} 
\usepackage{epstopdf}
\usepackage{float}
\usepackage{subcaption} 
\usepackage{booktabs}
\usepackage{amsthm} 
    
    \newtheorem{assumption}{Assumption}
    
    \newtheorem{property}{Property}
    \newtheorem{remark}{Remark}
    \newtheorem{theorem}{Theorem}

    \newtheorem{proposition}{Proposition}

\usepackage{fancyhdr}
\newcommand {\bv} {\left[\begin {array} {c} }
\newcommand {\ev} {\end{array}\right]}
\newcommand {\bmat} {\left[\begin{array} }
\newcommand {\emat} {\end{array}\right]}
\newcommand{\blista}{\renewcommand{\labelenumi}{(\roman{enumi})}
\begin{enumerate}}

\newcommand{\elista}{\end{enumerate} \renewcommand{\labelenumi}{\arabic{enumi}.}}
\newcommand {\beqn}{\begin{equation}}
\newcommand {\eeqn}{\end{equation}}
\newcommand {\beqna}{\begin{eqnarray}}
\newcommand {\eeqna}{\end{eqnarray}}
\newcommand {\beqnan}{\begin{eqnarray*}}
\newcommand {\eeqnan}{\end{eqnarray*}}

\newcommand{\numeq}[2][=]{\stackrel{\scriptstyle\mkern-1.5mu#2\mkern-1.5mu}{#1}} 
\def\moveEq#1{{}\mkern#1mu} 

\def\R{\mathbb{R}} 
\def\Z{\mathbb{Z}} 

\newcommand {\T}{^{\top}} 

\def\vv#1{\mathrm{\mathbf{#1}}} 
\newcommand{\vx} {\vv{x}}
\newcommand{\vu} {\vv{u}}

\newcommand{\tP}{\tilde{P}}

\newcommand{\setX}{\mathcal{X}}
\newcommand{\setU}{\mathcal{U}}
\newcommand{\setZ}{\mathcal{Z}}
\newcommand{\setW}{\mathcal{W}}
\newcommand{\setA}{\mathcal{A}}
\newcommand{\setB}{\mathcal{B}}
\newcommand{\setC}{\mathcal{C}}
\newcommand{\setD}{\mathcal{D}}
\newcommand{\setE}{\mathcal{E}}
\newcommand{\setL}{\mathcal{L}}
\newcommand{\setH}{\mathcal{H}}
\newcommand{\setV}{\mathcal{V}}
\newcommand{\setP}{\mathcal{P}}

\newcommand{\setK}{\mathcal{K}}

\newcommand{\bx}{\bar x}
\newcommand{\bu}{\bar u}
\def\tS{\tilde{S}}
\def\tY{\tilde{Y}}

\begin{document}
\begin{frontmatter}
\pagestyle{fancy}

\title{\LARGE \bf Tractable robust MPC design based on nominal predictions}

\author{Ignacio~Alvarado\corref{cor}}
\ead{ialvarado@us.es}

\author{Pablo~Krupa}
\ead{pkrupa@us.es}

\author{Daniel~Limon}
\ead{dlm@us.es}

\author{Teodoro~Alamo}
\ead{talamo@us.es}

\cortext[cor]{Corresponding author}

\address{Department of Systems Engineering and Automation, Universidad de Sevilla, Seville, Spain.}

\thispagestyle{fancy}

\begin{abstract}
Many popular approaches in the field of robust model predictive control (MPC) are based on nominal predictions.
This paper presents a novel formulation of this class of controller with proven input-to-state stability and robust constraint satisfaction.
Its advantages are: (i) the design of its main ingredients are tractable for medium to large-sized systems, (ii) the terminal set does not need to be robust with respect to all the possible system uncertainties, but only for a reduced set that can be made arbitrarily small, thus facilitating its design and implementation, (iii) under certain conditions the terminal set can be taken as a positive invariant set of the nominal system, allowing us to use a terminal equality constraint, which facilitates its application to large-scale systems, and (iv) the complexity of its optimization problem is comparable to the non-robust MPC variant.
We show numerical closed-loop results of its application to a multivariable chemical plant and compare it against other robust MPC formulations.
\end{abstract}


\thispagestyle{fancy}
\end{frontmatter}
\thispagestyle{fancy}
\pagestyle{fancy}
\section{Introduction} \label{sec:introduction}

A relevant problem that has received a lot of attention from the predictive control community is the robust regulation of disturbed linear discrete-time systems towards a desired equilibrium point.
However, in spite of the potential benefits of this paradigm, its implementation in the industry is deterred by their typical high complexity, especially for medium to large-scale systems \cite{Mayne_ARC_16}.

The classical approach for robust control laws, introduced in \cite{Wit68}, is to minimize a cost function for the worst possible disturbance realization. While this may lead to an optimal robust control law, it requires solving min-max optimization problems \cite{Bemporad_TAC_03}, \cite{SM98}, \cite{alamo2005constrained}, which may be very computationally demanding even for average-sized systems.

In order to overcome this, robust model predictive controllers (RMPC) \cite{Camacho_S_2013} based on nominal predictions and tightened constraints, typically referred to as \textit{tube-based} MPC, have been proposed in the literature.
As discussed in \cite{Zanon_ECC_2021}, ``the many variants [of tube-based MPC] developed over the years can essentially be classified as variations of two approaches [...] \cite{ChisciAUT01} and \cite{MayneAUT05}".
Other formulations, variations and approaches for RMPC have been proposed since then, such as non-linear RMPC formulations \cite{Kohler_ACC_2018}; \cite{Villanueva_AUT_2017}, which uses ellipsoidal robust forward invariant sets, resulting in an optimization problem with linear matrix inequalities (LMI); formulations which combine tube-based and multi-stage MPC such as \cite{Subramanian_Wiley_2021}; or self-triggered tube-based MPC \cite{Brunner_AUT_2016}.
However, we center our attention on the formulations from \cite{ChisciAUT01} and \cite{MayneAUT05}, since this paper proposes a similar formulation that provides several benefits over these two main tube-based RMPC formulations.

In \cite{ChisciAUT01}, the authors propose a RMPC controller where the effect of the disturbance is rejected with the use of a control gain $F$ taken from the solution of the linear quadratic regulator (LQR) corresponding to the weighting matrices of the cost function.
The system constraints are tightened by mean of \textit{reachable sets} which are computable for medium to large-scale systems (see \cite[Eq. (7)]{ChisciAUT01}).
The disadvantage is that $F$ is derived from the weighting matrices of the cost function. Thus, its effect on the constraint tightening \cite[Eq. (22)]{ChisciAUT01} cannot be freely tuned independently from the performance of the controller.

In \cite{MayneAUT05}, the authors propose a RMPC controller where the constraints are tightened using the minimal robust positive invariant set of the system, or an approximation of it \cite{RakovicTAC05}, for a given stabilizing control gain. This set is typically computationally demanding to obtain (in many cases prohibitively so), even for average-sized systems. The advantage is that the disturbance rejection (i.e., the constraint tightening) and the performance of the controller are decoupled, thus potentially leading to better performance.
This approach has been successfully applied to several applications, such as reference tracking \cite{LimonJPC10} and distributed control \cite{TroddenACC06}.

As previously mentioned, other similar constraint tightening approaches have been proposed. In \cite{RezaTAC12}, an initial feasible trajectory is calculated, and in the following sample times a control gain that keeps the system close to this initial trajectory is computed. This avoids having to solve an optimization problem online, since the deviation from the feasible trajectory is corrected using the linear feedback control gain. As such, this approach has a low computational burden; but in general provides worse performance than other robust MPC approaches \cite{GoulartAUT2006}.
In \cite{Subramanian_ACC_2017} the state estimation error is explicitly taken into consideration to provide a formulation that is independent of the  employed estimation method.
Finally, \cite{RossiterIJC08} proposes a RMPC for constrained linear systems described by polytopic uncertainty models. This approach significantly enlarges the feasibility region for small control horizons by introducing a new set of variables that have to be computed offline, thus increasing the complexity of the controller.

In this paper, we present a novel RMPC formulation based on nominal predictions and constraint tightening, guaranteeing input-to-state stability (ISS) (see \cite[\S 3]{LimonLNCIS09} for a definition of ISS) and robust satisfaction of the constraints.
Similarly to \cite{ChisciAUT01}, a robust control gain is used to tighten the constraints throughout the prediction horizon by taking into account the possible effect of the disturbances on the resulting closed-loop system.
However, this gain does not have to be the one corresponding to the LQR, as in \cite{ChisciAUT01}.
Instead, it can be freely tuned to enlarge the domain of attraction.
Moreover, the terminal set does not need to be robustly invariant for all the possible disturbances.
Instead, it only needs to be robust for a reduced set of disturbances within a certain set $\setL(N)$ whose size diminishes with the length of the prediction horizon $N$ of the controller, thus potentially leading to a larger terminal set than in other RMPC formulations.
An additional advantage of this is that if the prediction horizon is long enough to make the size of $\setL(N)$ negligible, then a positive invariant set of the nominal system (i.e., one that does not take into account the disturbance) can be used as the terminal set, significantly simplifying the design of the controller.
In fact, in this case a terminal equality constraint could be used, which would make the proposed controller applicable to large-scale systems.

The key points of the proposed formulation are:
\begin{itemize}
    \setlength\itemsep{-0.2em}
    \item The use of two decoupled design parameters (the constraint tightening robust control law, on one hand, and the terminal set, on the other), allows for a more flexible design of the controller, thus allowing for more opportunity to improve its performance.
    \item The tightened constraints share the same complexity as the nominal ones. That is, if the nominal constraints are box constraints, then the tightened ones are also box constraints. Moreover, they do not require the computation of the minimal robust positive invariant set.
    \item The design procedure seeks a robust control gain such that $\setL(N)$ decreases rapidly with $N$, thus increasing the size of the robust terminal set and the likelihood of being able to use a positive invariant set of the nominal system as the terminal set for reasonable values of $N$.
    \item We present design procedures for the ingredients of the controller that are tractable for average-sized systems, since they require solving optimization problems subject to LMI constraints.
    \item Its decision variables are the same as the ones of the nominal MPC variant.
        This, in addition with the previous points, results in an optimization problem with a complexity similar to that of nominal MPC (as is typical in tube-based RMPCs \cite{Mayne_AUT_2006}).
\end{itemize}

Given its computationally tractable design procedure and the fact that its resulting optimization problem is not particularly complex, we argue that the proposed approach simplifies the design of the controller, when compared with other robust predictive controllers, and may be applicable to many average-sized systems. To illustrate this, we show the results of controlling a simulated 12-state, 6-input chemical plant.

The remainder of this paper is structured as follows.
The problem statement is described in Section \ref{sec:problem}.
The proposed RMPC controller is detailed in Section \ref{sec:RMPC}.
We present design procedures for the computation of its main ingredients in Section \ref{sec:synthesis}.
Section \ref{sec:practical} discusses the use of a positive invariant set of the nominal system as the terminal constraint.
Section \ref{sec:case:study} shows two case studies, one controlling the chemical plant and one comparing the proposed formulations with \cite{ChisciAUT01} and \cite{MayneAUT05}.
We conclude with Section \ref{sec:conclusions}.


\vspace{0.5em}
\noindent{\textbf{Notation:} Given matrices $T$ and $P$, $T {\succ}{(\succeq)} 0$ indicates that $T$ is positive (semi)definite matrix and $T {\succ}{(\succeq)} P$ indicates $T - P {\succ}{(\succeq)} 0$.
For $x\in \R^{n}$ and $P \succ 0$, $\|x\| \doteq \sqrt{x\T x}$, $\|x\|_P \doteq \sqrt{ x\T P x}$, and $\| x \|_1 \doteq \sum\limits_{i=1}^{n} | x_{(i)} |$, where $x_{(i)}$ is the $i$-th component of $x$.
We denote by $(x_{1}, x_{2}, \dots, x_{N})$ the column vector formed by the concatenation of column vectors $x_{1}$ to $x_{N}$.
Given two integers $i$ and $j$ with ${j \geq i}$, $\Z_i^j$ denotes the set of integer numbers from $i$ to $j$, i.e. ${\Z_i^j \doteq \{i, i+1, \dots, j-1, j\}}$.
Given two sets $\setU \subset \R^n$ and $\setV \subset \R^n$, their Minkowski sum is defined by $\setU \oplus \setV \doteq \{u+v: \, u \in \setU,\, v \in \setV\}$, and their Pontryagin set difference is $\setU \ominus \setV \doteq \{u :\, u \oplus \setV \subseteq \setU\}$.
$I_n \in \R^{n \times n}$ denotes the identity matrix of dimension $n$.
For a symmetric matrix $M$, $\lambda_{\text{max}}(M)$ and $\lambda_{\text{min}}(M)$ denote its maximum and minimum eigenvalues, respectively.
Given $P \succ 0 \in \R^{n \times n}$, we denote the ellipsoid $\setE(P) \doteq \{ x \in \R^n : x\T P x \leq 1\}$.
We define the mapping of a set $\setU \subset \R^n$ with matrix $M \in \R^{m \times n}$ as $M \setU \doteq \{M u : u \in \setU\}$.
A function $f:\R_{\geq 0}\rightarrow\R_{\geq 0}$ is of class $\setK$ if it is continuous, strictly increasing and $f(0) = 0$, and is of class $\setK_\infty$ if it is a $\setK$-function and $f(x) \rightarrow + \infty$ as $x \rightarrow + \infty$.
The unitary box of dimension $n$ is denoted by $\setB_n \doteq \{x \in \R^n : \max\limits_i |x_{(i)}| \leq  1\}$, where $x_{(i)}$ is the $i$-th component of $x$.
For a given sequence of sets $\{\setV(i)\}_{i = 1}^N$, $\bigoplus\limits_{i=1}^{N}\setV(i) \doteq \setV(1) \oplus \setV(2) \oplus \dots \oplus \setV(N)$.
Given scalars and/or matrices $M_1, \dots, M_N$ (not necessarily of the same dimensions), we denote by $\texttt{diag}(M_1, \dots, M_N)$ the block diagonal matrix formed by their diagonal concatenation.

\section{Problem statement} \label{sec:problem}

We consider a plant described by the following controllable uncertain discrete-time linear time-invariant state-space model
\begin{equation} \label{eq:model}
    x(k+1) = A x(k) + B u(k) + w(k),
\end{equation}
where $x(k)\in \R^n$, $u(k)\in \R^m$ and $w(k)\in\R^n$ are the state, input and disturbance of the system at sampling time $k$, respectively.
Additionally, we consider that the state and input trajectories must satisfy the constraints $x(k)\in \setX$ and $u(k)\in \setU$ for any possible disturbance $w(k) \in \setW$, where the sets $\setX$ and $\setU$ are compact (convex) polytopes
\begin{subequations} \label{eq:sets:XU}
\begin{align}
    \setX &= \{x \in \R^n: A_{x}x\leq b_{x}\}, \label{eq:set:X}\\
    \setU &= \{u \in \R^m: A_{u}u\leq b_{u}\}, \label{eq:set:U}
\end{align}
\end{subequations}
with $A_x \in \R^{p_x \times n}$, $A_u \in \R^{p_u \times m}$, and which we assume contain the origin in their interiors; and set $\setW$ is a (convex) zonotope, i.e., an affine mapping of the unitary box of a certain dimension $M$
\begin{equation} \label{eq:set:W}
    \setW = H_W \setB_M,
\end{equation}
where $H_W \in \R^{n \times M}$.
We note that we consider $\setW$ as a zonotope to simplify future developments.

The control objective is to regulate the system to a neighborhood of the origin while fulfilling the constraints for all possible disturbances.

\section{Proposed robust MPC} \label{sec:RMPC} 

For a given prediction horizon $N$, the proposed robust MPC (RMPC) control law for a given state $x$ is derived from the solution of the following convex optimization problem, which we label by $\setP_N(x)$,
\begin{subequations} \label{eq:RMPC}
\begin{align}  
    \moveEq{-4}   \setP_N(x) \, : \, \min\limits_{\bar \vu} \; &V_N(x, \bar \vu) \\
    s.t.\;& \, \bx(i+1)=A\bx(i)+B\bu(i), \; i \in \Z_0^{N-1} \label{eq:RMPC:model}\\
        & \, \bx(0)=x  \label{eq:RMPC:initial} \\
        & \, \bx(i) \in \setX \ominus \setH(i), \; i \in \Z_0^{N-1} \label{eq:RMPC:const:x} \\
        & \, \bu(i) \in \setU \ominus K\setH(i), \; i \in \Z_0^{N-1} \label{eq:RMPC:const:u}\\
        & \, \bx(N) \in \Omega_{K_t} \ominus \setL(N), \label{eq:RMPC:terminal}
\end{align}
\end{subequations}
where $\bar\vu = (\bar u(0), \dots, \bar u(N{-}1))$; the cost function is
\begin{equation} \label{eq:RMPC:cost_function}
    V_N(x,\bar\vu) = \sum\limits_{i=0}^{N-1} \Big( \|\bar x(i) \|^2_Q + \|\bar u(i)\|^2_R \Big) + \|\bar x(N)\|^2_P
\end{equation}
for the cost function matrices $Q$, $R$ and $P$, which penalize the deviation between the predicted nominal evolution of the plant, i.e. \eqref{eq:model} with $w(i) = 0$ for all $i$, and the origin throughout the prediction horizon $N$; the sets $\setH(i)$ and $\setL(i)$ for $i \geq 1$ are given by
\begin{equation} \label{eq:sets:HL}
\setH(i) = \bigoplus_{j = 0}^{i - 1} {A_K^j \setW}, \quad \setL(i) = A_K^{i-1} \setW,
\end{equation}
where $A_K \doteq A + B K$, $K \in \R^{m \times n}$ is a linear feedback control gain and $\setH(0)$, $\setL(0)$ are taken as the empty sets;
and $\Omega_{K_t}$ is a robust positive invariant set of the system with the terminal feedback control gain $K_t$ for the disturbances contained in $\setL(N)$ (see Assumption \ref{ass:RMPC}.(iv) below).

Constraints \eqref{eq:RMPC:const:x} and \eqref{eq:RMPC:const:u} request not only that the predicted states and inputs satisfy the constraints $\setX$ and $\setU$, respectively, but instead that they lie within \textit{tightened constraints} that depend on the sets $\setH(i)$ and, so, on the feedback control gain $K$.
As shown in Appendix \ref{app:RMPC:feasibility:proof}, set $\setL(i)$ is a bound of the possible deviation at time instant $i$ that a disturbance $w \in \setW$ at the initial time instant, i.e. at $i = 0$, can create between model \eqref{eq:model} and the nominal model \eqref{eq:RMPC:model} if the control law $u = K(x - \bx) + \bu$ is used to reject it.
Notice that, if $A + B K$ is Hurwitz, the size of $\setL(i)$ monotonically decreases with $i$, whereas sets $\setH(i)$ monotonically increase, converging to a bounded set (the minimum robust positive invariant set).

\begin{assumption} \label{ass:RMPC}
    We make the following assumptions on the ingredients of optimization problem \eqref{eq:RMPC}:
\blista
    \item $Q, R \succ 0$.
    \item The control gain $K_t$ and $P \succ 0$ satisfy
        \begin{equation} \label{eq:ass:RMPC:P}
            P-(A+B K_t)^\top P(A+B K_t) \succeq Q+K_t^\top R K_t.
        \end{equation}
    \item The control gain $K$ is such that $A+BK$ is Hurwitz and $\setX \ominus \setH(N)$ and $\setU \ominus K\setH(N)$ are non-empty.
    \item The set $\Omega_{K_t}$ is a compact convex set satisfying
        \vspace*{-0.5em}
        \begin{align} 
            &\moveEq{-22}(A+BK_t)\Omega_{K_t} \oplus \setL(N) \subseteq \Omega_{K_t}, \label{eq:ass:RMPC:omega:invariant} \\
            &\moveEq{-22}\Omega_{K_t} \subseteq \{x \in \setX \ominus \setH(N) : K_t x \in \setU \ominus K \setH(N{-}1)\}. \label{eq:ass:RMPC:omega:admissible}
        \end{align}
\elista
\end{assumption}

One of the advantages of this formulation, when compared to other robust MPC approaches, such as \cite{ChisciAUT01}, is that it offers an extra degree of freedom. Indeed, the gain $K$, which is used to compute the sets $\setH(i)$ and $\setL(i)$, can be tuned in order to enlarge the region of attraction, whereas the gain $K_t$, which is used to compute the terminal cost function matrix $P$ and the set $\Omega_{K_t}$, is affected by the choice of $Q$ and $R$, which can be tuned to improve the performance of the controller (as is standard in MPC).
As stated in Assumption \ref{ass:RMPC}.(iv), $\Omega_{K_t}$ must be a robust positive invariant set of the system controlled with the terminal control law $K_t$ for the additive disturbances contained in $\setL(N) = A_K^{N-1} \setW$ (which is typically much smaller than $\setW$).
The prediction horizon $N$ can be chosen to obtain a small set $\setL(N)$, and therefore a larger terminal set and (generally) a larger domain of attraction of the controller.

In the following, we denote by $V_N^*(x)$ the optimal cost, $\bar \vu^*(x) = \{\bar u^*(0; x), \dots, \bar u^*(N-1; x)\}$ the optimal value of the decision variable and $\bar \vx^*(x) = \{\bar x^*(0; x), \dots, \bar x^*(N; x)\}$ the corresponding optimal value of the nominal state trajectory of problem $\setP_N(x)$. The control law at each sample time $k$ is given by the receding horizon control law $u(k) = \bar u^*(0;x(k))$, where $x(k)$ is the state of the plant.

The domain of attraction of the RMPC controller, denoted by $\setX_N$, is the feasibility region of $\setP_N(x)$, i.e., the set of states that can be steered to $\Omega_{K_t} {\ominus} \setL(N)$ in $N$ steps while fulfilling the tightened constraints \eqref{eq:RMPC:const:x} and \eqref{eq:RMPC:const:u}.

The following two theorems state the recursive feasibility and input-to-state stability of the RMPC controller for all initial states~${x \in \setX_N}$.

\begin{theorem}[Recursive feasibility] \label{theo:RMPC:feasibility}
    Consider a system \eqref{eq:model} as described in Section \ref{sec:problem} controlled with the robust MPC formulation $\setP_N(x)$. Suppose that the ingredients of the controller satisfy Assumption \ref{ass:RMPC} and that the system state at sample time $k$ satisfies $x(k) \in \setX_N$. Then, the successor state $x(k+1) = A x(k) + B \bar u^*(0; x(k)) + w(k)$ satisfies $x(k+1) \in \setX_N$ for any $w(k) \in \setW$.
\end{theorem}
\vspace*{-1.2em}
\begin{proof} \renewcommand{\qedsymbol}{}
    See Appendix \ref{app:RMPC:feasibility:proof}.
\end{proof}

\begin{theorem}[Input-to-state stability] \label{theo:RMPC:ISS}
    Consider a system \eqref{eq:model} as described in Section \ref{sec:problem} controlled with the robust MPC formulation $\setP_N(x)$. Suppose that the ingredients of the controller satisfy Assumption \ref{ass:RMPC} and that the system state at sample time $k$ satisfies $x(k) \in  \setX_N$. Then, the closed-loop system is ISS with respect to any disturbance signal $w(i) \in \setW$, $i \geq k$.
\end{theorem}
\vspace*{-1.2em}
\begin{proof} \renewcommand{\qedsymbol}{}
    See Appendix \ref{app:RMPC:ISS:proof}.
\end{proof}

\section{Synthesis of the RMPC ingredients} \label{sec:synthesis}

The proposed controller requires the design of the ancillary control gains $K$ and $K_t$, the matrix $P$, and the sets $\setH(i)$, $\setL(i)$, $\setX \ominus \setH(i)$, $\setU \ominus K \setH(i)$ and $\Omega_{K_t}$.
An appropriate design of these ingredients, which is not immediate, must ensure the satisfaction of the stabilizing assumptions (Assumption \ref{ass:RMPC}), reject the effect of the uncertainty and seek to maximizing the domain of attraction.

This section describes tractable procedures for the computation of these ingredients satisfying the stability conditions and guaranteeing robust constraint satisfaction whilst seeking to increase the domain of attraction. The procedures and results we show follow from prior results from the control literature, including \cite{LimonIWC08}, \cite{KothareAUT96}, \cite{Boyd_SIAM_1994_LMI} and \cite{Chen_ACC_2001}.
However, we present them here in a unified format and particularized to our proposed formulation.

\subsection{Computation of $K$} \label{sec:synthesis:K}

The control gain $K$ is used to compensate the deviation from the nominal predictions due to the disturbances. In this paper, we follow the approach from \cite{LimonIWC08}, in which the robustness criterium is to find control gain $K$ and a matrix $\tP \succ 0$ such that the ellipsoid $\setE(\tP)$ is a robust positive invariant set of system \eqref{eq:model} for the state feedback control law $u = K x$ satisfying the constraints $\setX$ and $\setU$. Additionally, we require the sets $\setX \ominus \setH(i)$ and $\setU \ominus K \setH(i)$ to be non-empty.
Moreover, we wish to minimize the size of $\setE(\tP)$. Note that, since $\setE(\tP)$ is a robust positive invariant set of the system, and $\setH(i)$ is contained in the minimum robust positive invariant set, a reduction of the size of $\setE(\tP)$ (typically) leads to a reduction of the sets $\setH(i)$, which is desirable since this translates into an enlargement of the tightened constraints. However, this reduction may come at the cost of increasing the gain of $K$, which may result in a reduction of the constraints $\setU \ominus K \setH(i)$. To avoid this, we impose that $K x \subseteq \rho \, \setU$, $\forall x \in \setE(\tP)$, where the role of the scalar $\rho$ satisfying $0 < \rho \leq 1$ is to limit the control action of $K$ in order to ensure a certain control authority.

The computation of $K$ and $\tP$ satisfying these criteria was posed in \cite[\S 4.3]{LimonIWC08} as an optimization problem involving LMI constraints (see also \cite{KothareAUT96}).
In the following, we detail how this optimization problem and LMI restrictions are posed, where in this paper we include an additional requirement with the objective of obtaining a control gain $K$ that provides a fast convergence to the origin of the autonomous system $x(k+1) = (A + B K) x(k) = A_K x(k)$.
The reason for doing so is to obtain a matrix $K$ such that the size of $\setL(N)$ quickly decreases with $N$, thus leading to a larger terminal set and domain of attraction for reasonable values of $N$.

Robust positive invariance of $\setE(\tP)$ can formulated as
\begin{equation} \label{eq:synthesis:K:cond:invariant}
    (A_K x {+} w)\T \tP (A_K x {+} w) \leq 1, \, \forall x \in \setE(\tP), \forall w \in \setW,
\end{equation}
which, considering the convexity with respect to $w$, only needs to be checked for all $w \in \text{vert}(\setW)$, where $\text{vert}(\setW)$ denotes the vertexes of $\setW$.
Applying the S-procedure, we have that the implication
\begin{equation*}
    x\T \tP x \leq 1 \Rightarrow (A_K x + w)\T \tP (A_k x + w) \leq 1
\end{equation*}
is satisfied if there exists a scalar $\lambda \geq 0$ such that
\begin{equation*}
    (A_K x {+} w)\T \tP (A_K x {+} w) + \lambda (1 - x\T \tP x) < 1,
\end{equation*}
which can be expressed as
\begin{equation*}
    \bmat{c} x \\ 1 \emat\T \bmat{cc} \lambda \tP - A_K\T \tP A_K & -A_K\T \tP w \\ -w\T \tP A_K & 1 - \lambda - w\T \tP w \emat \bmat{c} x \\ 1 \emat {>} 0.
\end{equation*}
Therefore, \eqref{eq:synthesis:K:cond:invariant} is satisfied if there exists $\lambda \geq 0$ such that
\begin{equation*}
    \bmat{cc} \lambda \tP - A_K\T \tP A_K & -A_K\T \tP w \\ -w\T \tP A_K & 1 - \lambda - w\T \tP w \emat \succ 0, \, \forall w {\in} \text{vert}(\setW).
\end{equation*}
This expression can be rewritten as
\begin{equation*}
    \bmat{cc} \lambda \tP & 0 \\ 0 & 1-\lambda \emat - \bmat{c} A_K\T \\ w\T \emat \tP \left[ A_K \; w \right] \succ 0,\, \forall w {\in} \text{vert}(\setW),
\end{equation*}
which applying the Schur complement, leads to
\begin{equation*}
    \bmat{ccc} \lambda \tP & 0 & A_K\T \\ 0 & 1 - \lambda & w\T \\ A_K & w & \tP^{-1} \emat \succ 0, \, \forall w {\in} \text{vert}(\setW).
\end{equation*}
Finally, by pre- and post-multiplying by $\texttt{diag}(\tP^{-1}, 1, I_n)$
and taking the transformations $S \doteq \tP^{-1}$ and $Y \doteq K \tP^{-1}$, we obtain the LMIs
\begin{equation}
\bmat{ccc} \lambda S & 0 & S A\T {+} Y\T B\T \\ 0 & 1 {-} \lambda & w\T \\ A S {+} B Y & w & S \emat {\succ} 0, \, \forall w {\in} \text{vert}(\setW). \label{eq:synthesis:K:cond:invariant:LMI}\\
\end{equation}

For all $x \in \setE(\tP)$, the state constraints $x \in \setX$ must be satisfied, i.e., $A_x x \leq b_x$, $\forall x \in \setE(\tP)$.
It is well known that $\max_{x \in \setE(\tP)} c\T x = \sqrt{c\T \tP^{-1} c}$. Therefore, the previous condition can be posed as
\begin{equation} \label{eq:synthesis:K:X:init}
    A_{x,j} \tP^{-1} A_{x,j}\T \leq b_{x,j}^2, \; j \in \Z_1^{p_x},
\end{equation}
where $A_{x,j}$ and $b_{x, j}$ are the $j$-th row/component of $A_x$ and $b_x$, respectively.
However, since we are interested in minimizing the size of $\setE(\tP)$, we impose the condition $\setE(\tP) \subseteq \sqrt{\gamma} \setX$, where admissibility of the solution requires that the scalar $\gamma$ satisfy $0 < \gamma \leq 1$. This condition can be imposed by adding $\gamma$ to the previous inequality, leading to
\begin{equation*}
    A_{x,j} \tP^{-1} A_{x,j}\T \leq \gamma b_{x,j}^2, \; j \in \Z_1^{p_x}.
\end{equation*}
Using the definition of $S$, this can be expressed as the LMIs
\begin{equation} \label{eq:synthesis:K:cond:X}
    A_{x,j} S A_{x,j}\T \leq \gamma b_{x,j}^2, \; j \in \Z_1^{p_x}.
\end{equation}

For all $x \in \setE(\tP)$, the input constraints $u = K x \in \setU$ must be satisfied, i.e, $A_u K x \leq b_u$, $\forall x \in \setE(\tP)$.
However, as discussed at the beginning of this subsection, we instead impose $K x \subseteq \rho \, \setU$, $\forall x \in \setE(\tP)$, where the scalar $\rho$ must satisfy $0 < \rho \leq 1$. Therefore, we impose $A_u K x \leq \rho b_u$, $\forall x \in \setE(\tP)$, which, once again, can be posed as 
\begin{equation*}
    A_{u,j} K \tP^{-1} K\T A_{u,j}\T \leq (\rho b_{u,j})^2, \, j \in \Z_1^{p_u}.
\end{equation*}
Then, from the definition of $S$ and $Y$, and applying the Schur complement, we have
\begin{align}
    &A_{u,j} K \tP^{-1} \tP \tP^{-1} K\T A_{u,j}\T \leq (\rho b_{u,j})^2, \, j \in \Z_1^{p_u}, \nonumber \\
    &A_{u,j} Y S^{-1} Y\T A_{u,j}\T \leq (\rho b_{u,j})^2, \, j \in \Z_1^{p_u}, \nonumber \\
    &\bmat{cc} (\rho b_{u, j})^2 & A_{u,j} Y \\ Y\T A_{u,j}\T & S \emat \succ 0, \, j \in \Z_1^{p_u}. \label{eq:synthesis:K:cond:U}
\end{align}

Finally, we want to find a matrix $K$ such that the autonomous system $x(k+1) = (A + B K) x(k)$ has a fast convergence to the origin.
To do this, we impose the following condition, where the contraction factor $\mu$ is a scalar selected in the range $0 < \mu \leq 1$,
\begin{equation*} \label{eq:synthesis:LMI:constraints}
    \mu \tP \succ A_K\T \tP A_K,
\end{equation*}
which following similar procedures leads to
\begin{align} \label{eq:synthes:K:cond:contraction}
    &\mu \tP^{-1} \tP \tP^{-1} - \tP^{-1} A_K\T \tP A_K \tP^{-1} \succ 0, \nonumber \\
    &\mu S - (S A\T + Y\T B\T) S^{-1} (A S + B Y) \succ 0, \nonumber \\
    &\bmat{cc} \mu S & S A\T + Y\T B\T \\ A S + B Y & S \emat \succ 0. 
\end{align}

Matrices $K$ and $\tP$ satisfying the above criteria can be recovered from the solution of the following optimization problem involving the LMIs \eqref{eq:synthesis:K:cond:invariant:LMI}, \eqref{eq:synthesis:K:cond:X}, \eqref{eq:synthesis:K:cond:U}, and \eqref{eq:synthes:K:cond:contraction},
\begin{equation} \label{eq:synthesis:K:LMI}
\begin{aligned} 
    &\min\limits_{Y, S, \gamma} \quad \gamma \\
    &s.t. \; \eqref{eq:synthesis:K:cond:invariant:LMI},\; \eqref{eq:synthesis:K:cond:X},\; \eqref{eq:synthesis:K:cond:U}, \;\eqref{eq:synthes:K:cond:contraction}.
\end{aligned}
\end{equation}
The procedure is to select values of $\rho$ and $\mu$, and to then solve the resulting problem \eqref{eq:synthesis:K:LMI} for increasing values of $\lambda$ until a feasible problem is found. If no feasible solution is found, then less restrictive values of $\rho$ and/or $\mu$ should be selected. An easy way to do this is to first fix $\mu = 1$ and reduce $\rho$, and to then fix $\rho$ and reduce $\mu$.

\begin{remark} \label{rem:we:compute:RPIS}
We note that problem \eqref{eq:synthesis:K:LMI} is a convex optimization problem that is solved offline. Additionally, it can be solved for (relatively) large-sized systems guaranteeing a good design of the controller, although we remark that there is no guarantee that the problem will be feasible, since there may not exist a $K$ for which the tightened constraints are non-empty if $\setW$ is too large.
\end{remark}

\begin{remark} \label{rem:synthesis:K:numerical:issues}
    We note that elements $b_{u, j}$ and $b_{x, j}$ from \eqref{eq:synthesis:K:cond:X} and \eqref{eq:synthesis:K:cond:U} can cause numerical issues when solving problem \eqref{eq:synthesis:K:LMI}, since they appear squared in the LMI constraints. To avoid this, sets $\setX$ and $\setU$ should be rewritten so that the components of $b_x$ and $b_u$ only contain the value $1$, which is possible since we assume that the origin is an interior point of $\setX$ and $\setU$.
\end{remark}

\subsection{Computation of the tightened constraints} \label{sec:synthesis:constraints}

The computation of the tightened constraints requires the computation of sets $\setH(i)$ and $\setL(N)$ \eqref{eq:sets:HL}.
Set $\setL(N)$ is straightforward, since it is the zonotope given by
\begin{equation*}
    \setL(N) = A_K^{N-1} H_W \setB_M,
\end{equation*}
and sets $\setH(i)$ can be computed recursively by making use of the following proposition.

\begin{proposition}[\cite{Kuhn1998zonotopes}, Lemma 2(3)] \label{prop:Minkowsky:zonotope}
Let $\mathcal{C} \doteq H_C \setB_{M_C}$, with $H_C \in \R^{n \times M_C}$, and $\mathcal{D} \doteq H_D \setB_{M_D}$, with $H_D \in \R^{n \times M_D}$, be two zonotopes. Then, $\mathcal{C} \oplus \mathcal{D} = \left[ H_C, \; H_D \right] \setB_{M_C + M_D}$.
\end{proposition}

Indeed, $\setH(i) = H_{i} \setB_{iM}$, where $H_1 = H_W$ and
\begin{equation*}
    H_i = [A_K^{i-1}H_W, \; H_{i-1}], \; i > 1.
\end{equation*}

The tightened constraints $\setX \ominus \setH(i)$ and $\setU \ominus K \setH(i)$ are easily computed by using the following well-known result, for which we include a proof for completeness.

\begin{proposition} \label{prop:Pontryagin:zonotope}
    Let ${D \doteq H_D \setB_M}$, with $H_D \in \R^{n \times M}$, and let $\setC \doteq \{ x \in \R^n : F x \leq f \}$. Then, ${\setZ \doteq \setC \ominus \setD}$ is given by, $\setZ = \{ x \in \R^n : F z \leq f - g \}$,
where, denoting $F_i$ the $i$-th row of $F$, each component $i$ of $g$ is given by $g_i = \|F_i H_D\|_1$.
\end{proposition}

\begin{proof}
    $z \in \setZ$ if $x = z + d \in \setX$ for all $d \in \setD$. This can be posed as $F (z + d) \leq f$ for all $d \in \setD$. From the definition of set $\setD$, this is equivalent to $F z + F H_D v \leq f$ for all $v \in \setB_M$. The most restrictive value of $v$ for each linear inequality in $F$ is given by $\max_{v \in \setB_M} F_i H_D v = \| F_i H_D \|_1$. 
\end{proof}

\begin{remark} \label{rem:tightened:constraint:complexity}
    Note that Proposition \ref{prop:Pontryagin:zonotope} shows that the tightened constraints $\setX \ominus \setH(i)$ and $\setU \ominus K \setH(i)$ have the same complexity as the nominal ones \eqref{eq:sets:XU}, since they are polytopes with the same matrices $A_x$ and $A_u$.
    Additionally, since the computation of the sets are done offline and they only require vector norms and vector-matrix multiplications, this procedure can be applied to large-scale systems.
\end{remark}

\subsection{Computation of $K_\MakeLowercase{t}$, $P$ and $\Omega_{K_\MakeLowercase{t}}$} \label{sec:synthesis:omega}

We follow a similar procedure to the one used in Section \ref{sec:synthesis:K}. That is, we compute matrices $K_t$ and $P \succ 0$ satisfying \eqref{eq:ass:RMPC:P}, and such that $\Omega_{K_t} \doteq \setE(P)$ satisfies \eqref{eq:ass:RMPC:omega:invariant} and \eqref{eq:ass:RMPC:omega:admissible}. As done in Section \ref{sec:synthesis:K}, these conditions can be imposed as LMIs as follows, where we define $\tS {\doteq} P^{-1}$ and $\tY {\doteq} K_t P^{-1}$.

\begin{remark} \label{rem:selection:Omega}
The design procedure that we detail below provides the values of $K_t$ and $P$.
Additionally, it shows the existence of a robust positive invariant set of the form $\Omega_{K_t} = \setE(P)$. 
The use of an ellipsoidal terminal invariant set is useful due to it typically being simpler to compute that the more common polyhedral invariant set \cite[\S  4.1]{Blanchini_A_1999}, \cite{Wan_AUT_03} and to it resulting in the addition of fewer constraints in the optimization problem \cite[\S 5]{Blanchini_A_1999}.
However, a polyhedral robust invariant set $\Omega_{K_t}$ can be computed by other means \cite{Fiacchini_AUT_2010}, \cite{Blanchini2008set}, and used instead of the ellipsoidal one obtained from the following design procedure.
\end{remark}

Condition \eqref{eq:ass:RMPC:P} can be posed as an LMI as follows,
\begin{align}
    &\tS {-} (\tS A\T {+} \tY\T B\T) P (A \tS {+} B \tY) \succ \tS Q \tS + \tY\T R \tY, \nonumber \\
    &\tS - \bmat{c} A \tS {+} B \tY \\ Q^{1/2} \tS \\ R^{1/2} \tY \emat\T \bmat{ccc} P & 0 & 0 \\ 0 & I & 0 \\ 0 & 0 & I \emat \bmat{c} A \tS {+} B \tY \\ Q^{1/2} \tS \\ R^{1/2} \tY \emat {\succ} 0, \nonumber \\
    &\bmat{cccc} \tS & * & * & * \\ A \tS + B \tY & \tS & * & * \\ Q^{1/2} \tS & 0 & I & * \\ R^{1/2} \tY & 0 & 0 & I \emat \succ 0, \label{eq:synthesis:omega:LQR}
\end{align}
where we use the asterisks to represent the transposed of the elements shown in the lower half of the matrix due to space considerations.
Condition \eqref{eq:ass:RMPC:omega:invariant} can be posed as
\begin{equation*}
    (A_K x {+} d)\T P (A_K x {+} d) \leq 1, \, \forall x \in \setE(\tP), \forall d \in \setL(N),
\end{equation*}
which, using the arguments used to derive \eqref{eq:synthesis:K:cond:invariant:LMI}, leads to
\begin{equation} \label{eq:synthesis:omega:invariant}
    \bmat{ccc} \tilde \lambda \tS & * & * \\ 0 & 1 {-} \tilde \lambda & * \\ A \tS {+} B \tY & d & \tS \emat \succ 0, \, \forall d {\in} \text{vert}(\setL(N)),
\end{equation}
for some $\tilde \lambda \geq 0$.

Finally, we pose condition \eqref{eq:ass:RMPC:omega:admissible} as two LMIs. Note that, from Section \ref{sec:synthesis:constraints} (see Remark \ref{rem:tightened:constraint:complexity}), we have that the tightened constraints $\setX \ominus \setH(N)$ and $\setU \ominus K \setH(N-1)$ are compact (convex) polytopes given by
\begin{align*}
    \setX \ominus \setH(N) &= \{x \in \R^n: A_{x}x\leq \tilde b_{x}\}, \\
    \setU \ominus K \setH(N-1) &= \{u \in \R^m: A_{u}u\leq \tilde b_{u}\},
\end{align*}
where $\tilde b_x$ and $\tilde b_u$ are computed as described Proposition \ref{prop:Pontryagin:zonotope}. Therefore, condition \eqref{eq:ass:RMPC:omega:admissible} can be posed as LMIs following the same procedure used to derive \eqref{eq:synthesis:K:X:init} and \eqref{eq:synthesis:K:cond:U}:
\begin{align}
    &A_{x,j} \tS A_{x,j}\T \leq \tilde b_{x,j}^2, \; j \in \Z_1^{p_x}, \label{eq:synthesis:omega:cond:X} \\
    &\bmat{cc} \tilde b_{u, j}^2 & A_{u,j} \tY \\ \tY\T A_{u,j}\T & \tS \emat \succ 0, \, j \in \Z_1^{p_u}, \label{eq:synthesis:omega:cond:U}
\end{align}
where we note that $\rho$ is not used in this case. Remark \ref{rem:synthesis:K:numerical:issues} also applies to the two above LMIs.

The computation of $K_t$, $P$ and $\Omega_{K_t}$ can therefore be recovered by finding a feasible solution of the LMIs \eqref{eq:synthesis:omega:LQR}, \eqref{eq:synthesis:omega:invariant}, \eqref{eq:synthesis:omega:cond:X} and \eqref{eq:synthesis:omega:cond:U} for some value of $\tilde \lambda \geq 0$.

\section{Using a positive invariant set of the nominal system as the terminal set} \label{sec:practical}

As discussed in \cite{Mayne_ARC_16}, the application of RMPC to real systems is hindered because of its complexity, particularly for medium to large-scale systems. A mayor contributor to this complexity comes from the use of a robust positive invariant terminal set, particularly in the case of a polyhedral one.

As shown in Assumption \ref{ass:RMPC}.(iv) and \eqref{eq:RMPC:terminal}, our proposed formulation requires a terminal set that must be robust for the disturbances contained in $\setL(N)$, which will typically satisfy $\setL(N) \subset \setW$.
This is by itself a useful property of the formulation, since it simplifies the computation of $\Omega_{K_t}$ and results in a larger terminal set that in other RMPC formulations.

However, note that if $\setL(N) = \{0\}$, then the problem is simplified further.
Indeed, first note that the terminal set in \eqref{eq:RMPC:terminal} reduces to $\Omega_{K_t}$, and, most importantly, that the conditions stated in Assumption \ref{ass:RMPC}.(iv) reduce to
\begin{subequations} \label{eq:omega:non:robust}
\begin{align}
            &(A+BK_t)\Omega_{K_t} \subseteq \Omega_{K_t}, \label{eq:omega:non:robust:invariant} \\
            &\Omega_{K_t} \subseteq \{x \in \setX \ominus \setH(N) : K_t x \in \setU \ominus K \setH(N{-}1)\}. \label{eq:omega:non:robust:constraints}
\end{align}
\end{subequations}
Thus, $\Omega_{K_t}$ has to be a positive invariant set for the system $x(k+1) = (A + B K_t) x(k)$ subject to the constraints $x(k) \in \setX \ominus \setH(N)$ and $K_t x(k) \in \setU \ominus K \setH(N-1)$.
That is, $\Omega_{K_t}$ must be a positive invariant set of the nominal (non-disturbed) system \eqref{eq:model} controlled with the state feedback gain $K_t$ for the constraints in \eqref{eq:omega:non:robust:constraints}.
This is a significant benefit over having to compute a robust positive invariant set, since we can take the positive invariant set used in nominal MPC formulations, which in general is much simpler to compute than one that has to be robust to some uncertainty.
Additionally, we can use very simple terminal sets, such as the ellipsoidal set $\Omega_{K_t} = \setE(P)$ obtained following the method in Section \ref{sec:synthesis:omega}, or a terminal equality constraints, since in this case the set $\Omega_{K_t} = \{0\}$ satisfies \eqref{eq:omega:non:robust}.
In these two cases the proposed formulation could be applied even to medium to large-scale systems.

The reader may note  that the condition $\setL(N) = \{0\}$ is impractical, since in general $\setL(N) \rightarrow \{0\}$ as $N \rightarrow +\infty$. If $K$ is taken as the gain of the dead-beat controller then $N$ can be selected such that $A_K^{N-1} = 0$. The problem with this approach is that, typically, the resulting matrix $K$ will have a big gain, and therefore the tightened constraints $\setU \ominus K \setH(i)$, $i \in \Z_0^{N-1}$, will be too restrictive (they might even be empty).
Therefore, instead of forcing $\setL(N) = \{0\}$, we propose to relax this condition to $\setL(N)$ being very small by forcing the norm of $A_K^{N-1}$ to be sufficiently small. In particular, we take the square of the spectral norm of $A_K^{N-1}$ (which is the absolute value of its maximum eigenvalue).
If this norm is smaller than a certain tolerance, then $\setL(N)$ can be taken as $\{0\}$ for all practical purposes.
Note that the design procedure we present in Section \ref{sec:synthesis:K} seeks a control gain $K$ that provides a fast convergence of the autonomous system $x(k+1) = A_K x(k)$.
Thus, it is expected to provide a $K$ such that a negligible $\setL(N)$ is attained for reasonable values of $N$.

A more practical approach is to relax this condition even further. The optimization problem of the RMPC controller will be solved, in real-time, using some iterative optimization algorithm with an exit condition determined by an exit tolerance set by the user. In practice, the exit tolerance is typically in the range of $10^{-3}$ to $10^{-6}$. Therefore, we argue that, in practice, it is enough to select $N$ and compute $K$ so that the norm of $A_K^{N-1}$ is significantly smaller than the exit tolerance of the solver. In this case, for all practical purposes, the terminal equality constraint $\bx(N) = 0$ can also be used.

\begin{remark} \label{rem:control:horizon}
The control gain $K$ computed as described in Section \ref{sec:synthesis:K} may still lead to a large $N$ in order for $\setL(N)$ to be sufficiently small. In this case, to reduce the complexity of optimization problem \eqref{eq:RMPC}, a control horizon $N_c$ may be used alongside the prediction horizon $N$, as is typically done in MPC, by adding constraint $\bu(i) = K_t \bx(i)$, $i \in \Z_{N_c}^{N-1}$, to~\eqref{eq:RMPC}.
Even if $N$ has to be large, in many cases the resulting optimization problem will still be implementable, since many solvers for linear MPC can exploit the structure of the optimization problem, leading to a linear memory and computational complexity growth with the prediction horizon $N$ \cite{Krupa_arXiv_ellipMPC_21}, \cite{Quirynen_OCAM_2020}, \cite{Domahidi_FORCES}.
\end{remark}

\begin{remark} \label{rem:practical:eliminate}
Another approach to deal with the issues that arise from the need to compute and implement a terminal set is to eliminate it altogether. This can be done by increasing the penalization of the terminal cost with an additional scalar penalty parameter $\lambda \geq 0$ \cite{LimonLNCIS09}, i.e. to take $\lambda \| \bx(N) \|^2_P$ in \eqref{eq:RMPC:cost_function}. However, the issue is that it is not clear how to select $\lambda$, since the region in which the controller remains robust is not known a priori \cite{LimonTAC06}.
\end{remark}

\begin{remark} \label{rem:synthesis:omega:equality}
    If the option of taking $\Omega_{K_t}$ as the singleton $\Omega_{K_t} = \{0\}$ is taken as discussed above, then the ingredients $K_t$ and $P$ only needs to satisfy \eqref{eq:ass:RMPC:P}.
    In this case, the most straightforward choice is to take \eqref{eq:ass:RMPC:P} as en equality, i.e., to solve the synthesis problem of the discrete LQP problem, as is standard in MPC.
    The reason for our choice of considering \eqref{eq:ass:RMPC:P} as an inequality is to provide a wider selection range for $K_t$ and $P$ if $\Omega_{K_t}$ is not taken as a singleton.
\end{remark}

\section{Case study}  \label{sec:case:study}

We show two case studies, one on a multivariable chemical plant, showing that the proposed formulation is applicable to medium-sized systems, and one to an academic example in which we compare the terminal sets and domains of attraction of the proposed formulation with the ones from \cite{ChisciAUT01} and \cite{MayneAUT05}.

\subsection{Multivariable chemical plant} \label{sec:case:study:plant}

\begin{figure}[t]
    \centering
    \includegraphics[width=1\columnwidth]{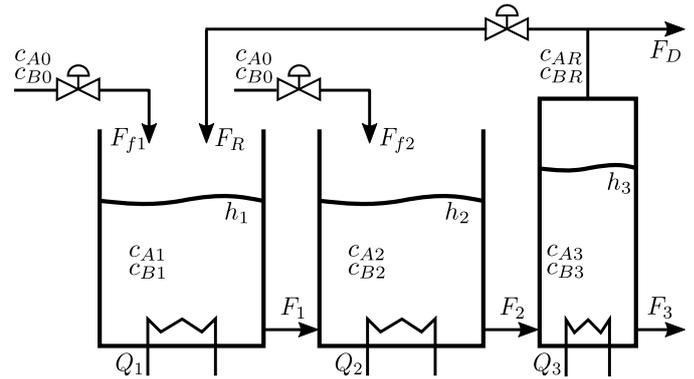}
    \caption{Double reactor and separator system.}
    \label{fig:Reactors}
\end{figure}

\begin{figure*}[t]
    \centering
    \begin{subfigure}[ht]{0.48\textwidth}
        \includegraphics[width=\linewidth]{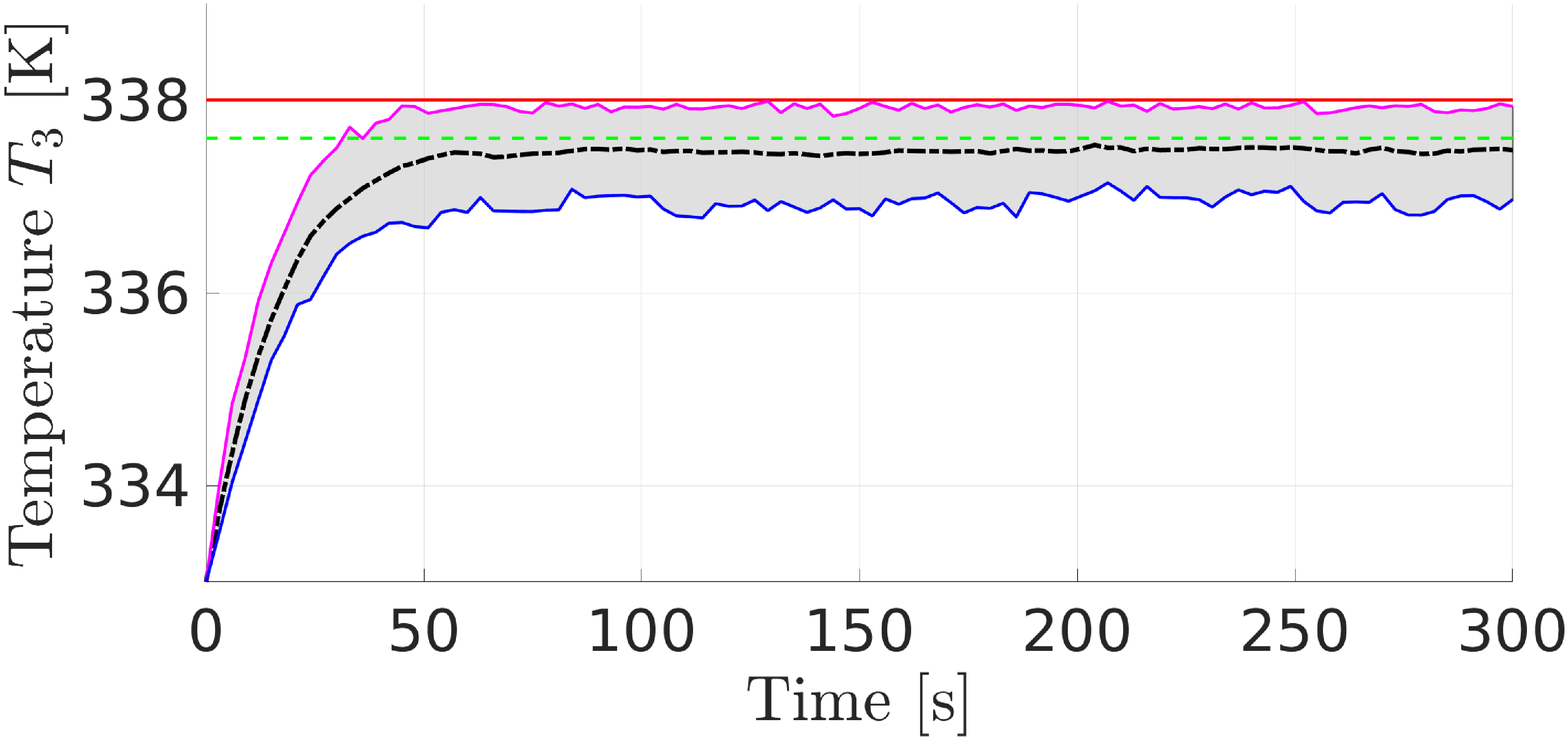}
        \caption{RMPC.}
        \label{fig:reactors:tests:r1:RMPC}
    \end{subfigure}%
    \quad
    \begin{subfigure}[ht]{0.48\textwidth}
        \includegraphics[width=\linewidth]{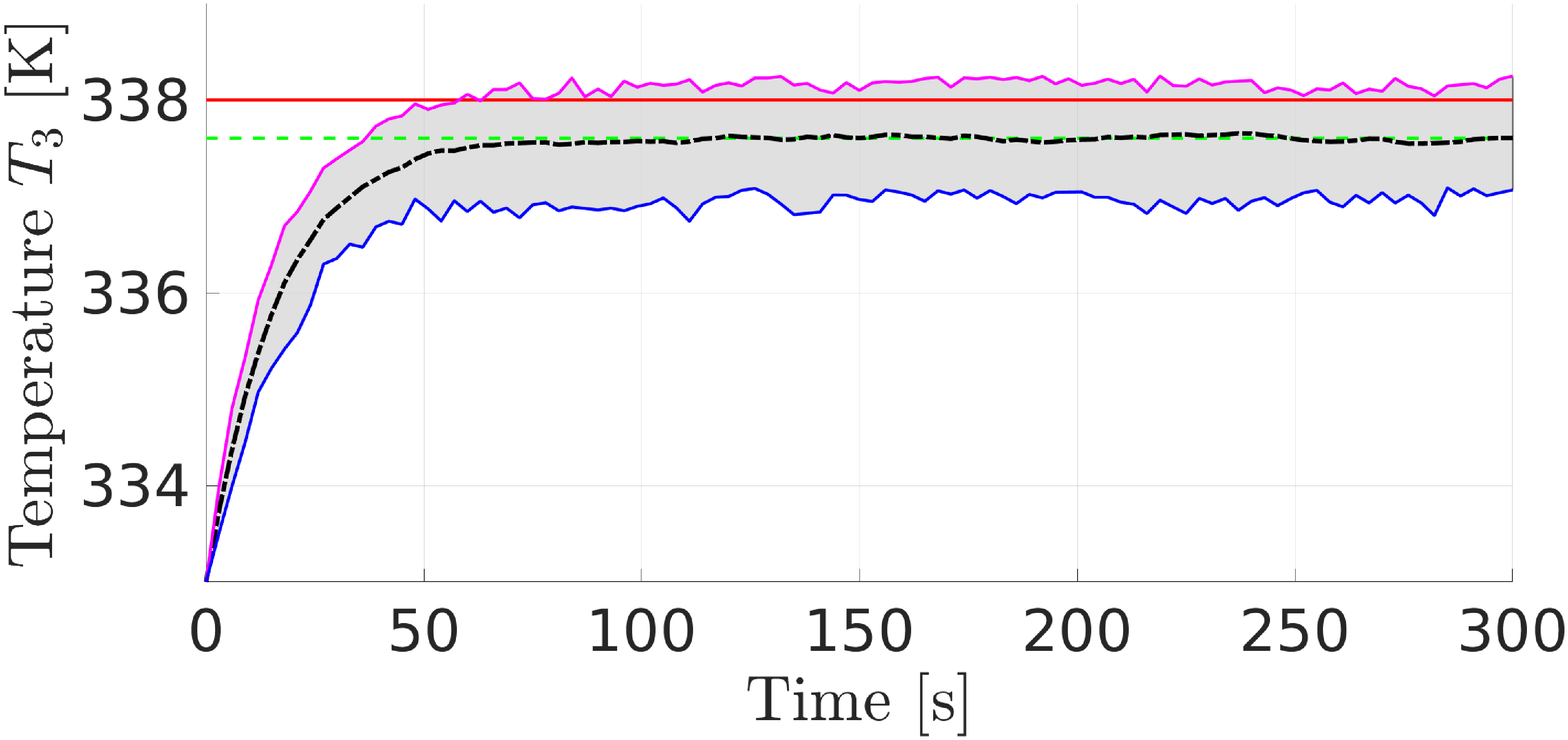}
        \caption{Nominal MPC.}
        \label{fig:reactors:tests:r1:MPC}
    \end{subfigure}%
    \caption{Closed-loop results of $T_3$ for the RMPC and nominal MPC controllers with the double reactor and separator plant.}
    \label{fig:reactors:tests}
\end{figure*}

\begin{figure}[t]
    \begin{center}
    \includegraphics[width=0.96\linewidth]{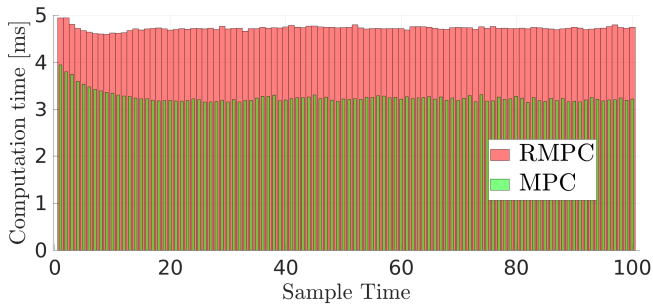}
    \caption{Computation times of the nominal and robust controllers.}
    \label{fig:times}
    \end{center}
\end{figure}

This section presents a case study where the multivariable chemical plant described in \cite[\S 5.7.1]{Krupa_Thesis_21} and depicted in Figure~\ref{fig:Reactors} is controlled using the proposed RMPC formulation. 
This plant is a 12-state, 6-input and 4-output system consisting of two consecutive reactors and a separator where two first-order reactions, ${A \rightarrow B}$ and ${B \rightarrow C}$, take place.
We take the parameters, constraints and operating point shown in \cite[\S 5.7.1]{Krupa_Thesis_21} and the reference as the one shown in \cite[Table 5.5]{Krupa_Thesis_21}.
We include disturbances $w$ acting on the heights and temperatures of the three volumes, given by a uniform distribution on the intervals $\pm0.01$ for the heights and $\pm0.25$ for the temperatures.
We obtain a model \eqref{eq:model} of the system by linearizing around the operating point and taking a sampling time of $3$s.

We design the RMPC controller following the procedures described in Section \ref{sec:synthesis} taking $Q = 5 I_n$, $R = 0.5 I_m$, $\rho = 1$ and $\mu = 0.9$.
Optimization problem \eqref{eq:synthesis:K:LMI} is constructed using the YALMIP package \cite{Lofberg2004} for Matlab, and solved using the SDPT3 solver \cite{SDPT3_99} for increasing values of $\lambda$.
A feasible solution is found for $\lambda = 0.701$, where the optimal value of $\gamma$ is $0.3358$. Taking $N = 60$, the square of the spectral norm of $A_k^N$ is $4.06 \cdot 10^{-6}$, which is relatively small.
Therefore, we consider a terminal equality constraint $\bx(N) = 0$, as discussed in Section \ref{sec:practical}, and compute $P$ by solving the LQR synthesis problem for our choice of $Q$ and $R$ (see Remark \ref{rem:synthesis:omega:equality}).

We use this model to simulate the system, since we are interested in determining if the proposed formulation does, indeed, robustly control model \eqref{eq:model} for our choice of $w$, as stated in Theorems \ref{theo:RMPC:feasibility} and \ref{theo:RMPC:ISS}.
To this end, we compare our proposed formulation with a nominal MPC controller using the same formulation and ingredients (including the terminal equality constraint), but that, obviously, considers the nominal constraints instead of the tightened ones.
Notice that the optimization problems of the RMPC and the nominal MPC are nearly identical, differing only in that the RMPC uses tightened state and input constraints.

\begin{figure*}[t]
    \centering
    \begin{subfigure}[ht]{0.48\textwidth}
        \includegraphics[width=\columnwidth, height=5cm]{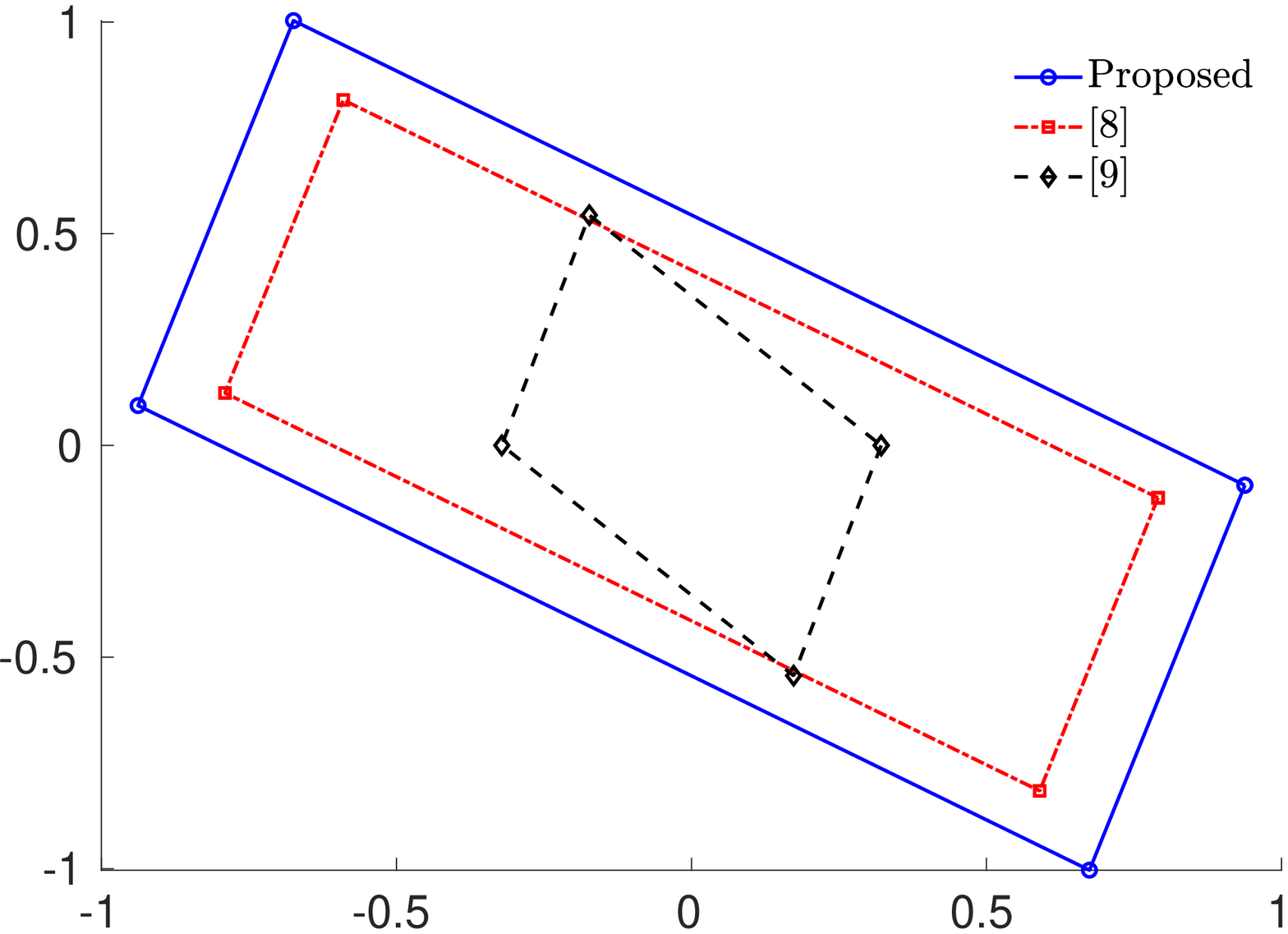}
        \caption{Terminal sets.}
        \label{fig:set:comparison:terminal}
    \end{subfigure}%
    \quad
    \begin{subfigure}[ht]{0.48\textwidth}
        \includegraphics[width=\columnwidth, height=5cm]{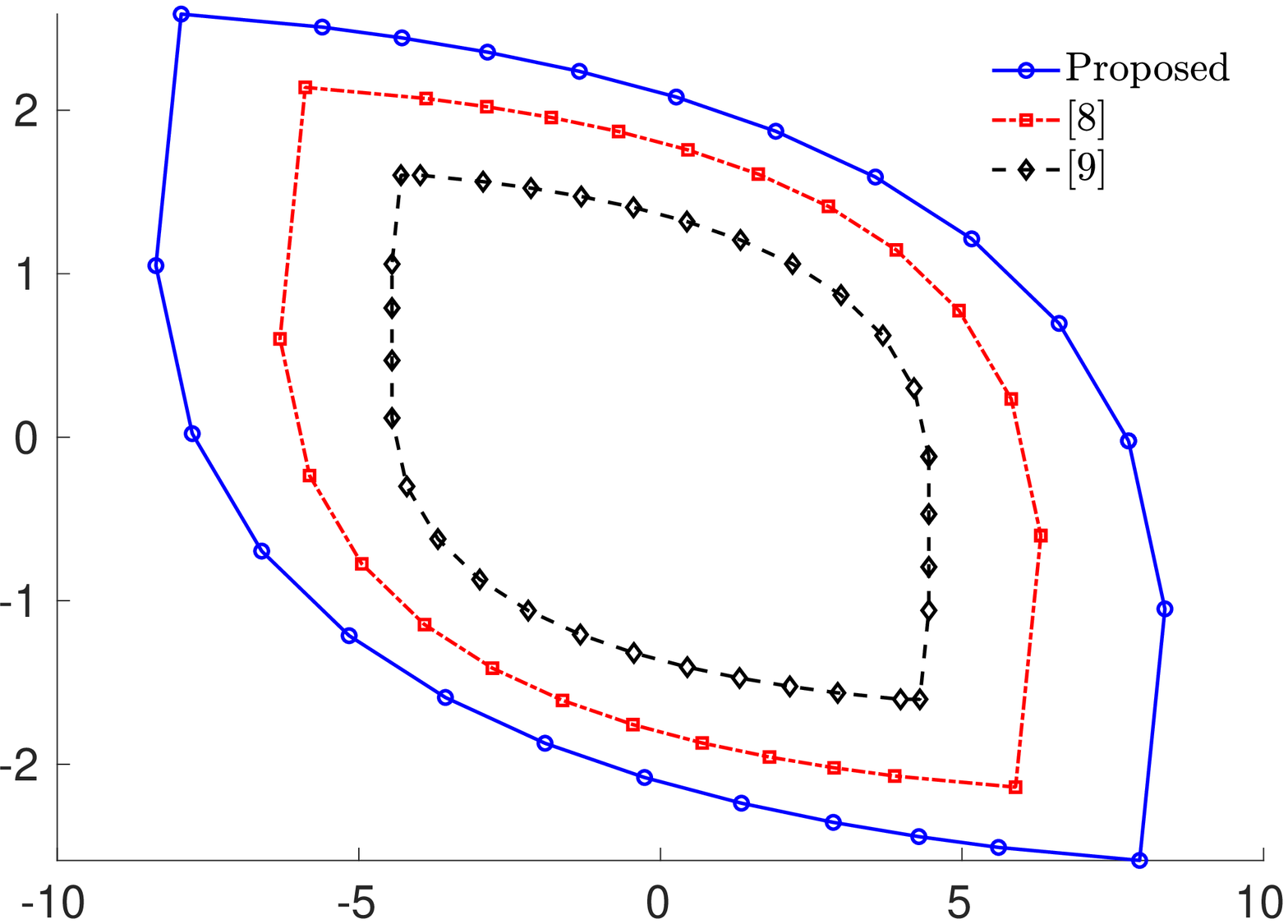}
        \caption{Domain of attraction.}
        \label{fig:set:comparison:domain}
    \end{subfigure}%
    \caption{Terminal set and domain of attraction of \cite{ChisciAUT01}, \cite{MayneAUT05} and \eqref{eq:RMPC}.}
    \label{fig:set:comparison}
\end{figure*}

Figure \ref{fig:reactors:tests} shows the trajectories of the temperature of the separator, $T_3$, using the RMPC and MPC formulations to control model \eqref{eq:model}.
Each figure shows the result of $100$ tests with different realizations of the disturbances, where the same realizations are used for both MPC controllers.
We highlight the maximum and minimum temperature attained at each iteration in magenta and blue, respectively; the average temperature of the tests at each sample time in dash/dotted black; the reference in dashed green; and the upper bound of $T_3$ in red.
As can be seen in the figures, the MPC controller sometimes violates the constraint, whereas the RMPC controller does not; at the expense of not being able to track the reference with zero offset.
For references sufficiently far away from the constraints both controllers have the same behaviour.
We focus on the evolution of $T_3$ since it is the only state that shows active (and violated) constraints during the simulations.
Appendix \ref{app:extended} shows the evolution of the other states and control inputs.

Both formulations are solved using the ADMM-based solver for the MPC formulation with terminal equality constraint presented in \cite{Krupa_TCST_20} from version \texttt{v0.3.4} of the SPCIES toolbox \cite{Spcies}, where we take the exit tolerance as $10^{-4}$ and the penalty parameter of the ADMM algorithm as $\rho = 15$.
Figure \ref{fig:times} shows the average computation times of both solvers at each sample time of the simulation using an Intel Core i5-8250U CPU operating at $1.60$ GHz.
The difference between the two is due to the fact that the RMPC solver tends to have more active constraints, which typically increase the number of iterations of first order methods.
The maximum and minimum computation times, in milliseconds, for each solver are $7.76$ and $4.43$ for the RMPC, and $7.49$ and $3.04$ for the MPC, respectively.

\subsection{Comparison with other RMPC formulations} \label{sec:case:study:comparison}

This section compares the terminal region and domain of attraction of the proposed formulation with the ones of the RMPC formulations from \cite{ChisciAUT01} and \cite{MayneAUT05}. 
We consider the example from \cite[\S 5]{ChisciAUT01} taking the constraints as $| x | \leq 10$, $| u | \leq 1$ and $| w | \leq 0.16$.
We take $Q = I_2$, $R = 0.01$ and $N = 10$ in the three RMPC formulations.
We use the LQR gain for these cost function matrices as the $K_t$ gain used to compute the set $\Omega_{K_t}$, which we take as the maximal polytopic robust invariant set.
We also use this gain to compute the tightening of the constraints and the terminal set of the formulation \cite{ChisciAUT01} as well as for the constraint tightening for the formulation \cite{MayneAUT05}.
We use the LQR gain for the matrices $Q = I_2$ and $R = 100$ to compute the tightened constraints of our proposed formulation.
For the formulation from \cite{MayneAUT05}, we take the LQR gain for the matrices $Q = \texttt{diag}(100, 0)$ and $R = 0.01$ to compute the minimal robust positive invariant set.
The gains for the proposed formulation and for the formulation from \cite{ChisciAUT01} were hand tuned to increase their domain of attraction and terminal sets, whereas the gain used to compute the minimal robust positive invariant set for \cite{MayneAUT05} was hand tuned to produce the smallest one possible, thus reducing the size of the associated tube.
All the sets are computed using the MPT3 toolbox \cite{MPT3}.

Figure \ref{fig:set:comparison} shows the comparison between the three formulations, where Figure \ref{fig:set:comparison:terminal} shows the terminal sets and Figure \ref{fig:set:comparison:domain} the domains of attraction.
As can be seen, the proposed formulation provides a larger terminal region and domain of attraction. This is partly due to the fact that we have two degrees of freedom, one for the tightened constraints ($K$) and one for the set $\Omega_{K_t}$, and partly due to the fact that the terminal set only has to be robust for a subset of the disturbances $\setW$, i.e., $\setL(N)$.

\section{Conclusions} \label{sec:conclusions}

This paper presents a robust MPC formulation based on nominal predictions that \textit{(i)} uses two independent control gains, one, derived from the MPC cost function matrices, related to the performance, and one related to the constraint tightening, thus providing a good trade-off between performance and domain of attraction, \textit{(ii)} the computation of the terminal set is simplified thanks to it not having to be robust for all the possible system uncertainties, but only for a reduced-sized set, and \textit{(iii)} is recursively feasible and stable in the ISS sense.
Additionally, we provide tractable procedures for the computation of its ingredients.
This, along with the possibility of being able to use a positive invariant set of the nominal system as the terminal set, i.e., for the system without the disturbance, results in a formulation that could be applied to relatively large-sized systems.
In particular, in this case the resulting optimization problem would share the same complexity as its equivalent nominal MPC formulation, although, as is typical in tube-based robust MPC formulations, the closed-loop performance of the proposed controller may be significantly worse than its nominal counterpart.

\begin{figure}[t]
    \begin{center}
\includegraphics[width=\columnwidth]{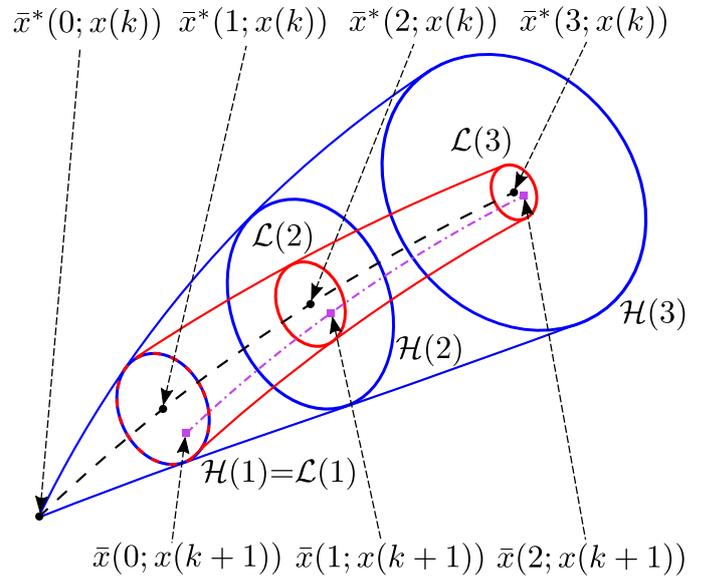}    
\caption{Visual representation of the feasible state trajectories. Black dashed line and dots represents the optimal trajectory at sample time $k$, magenta dashed-dotted line and squares the feasible trajectory for the successor state, blue ellipsoids the sets $\setH(i)$ and red ellipsoids the sets $\setL(i)$. Solid blue and red lines are for visual aid.}
    \label{fig:feasibility:trajectory}
    \end{center}
\end{figure}

\section*{Declaration of competing interest}

\noindent The authors declare that they have no known competing financial interests or personal relationships that could have appeared to influence the work reported in this paper.

\section*{Acknowledgments}

\noindent This work was supported in part by Grant PID2019-106212 RB-C41 funded by MCIN/AEI/10.13039/501100011033,
in part by Grant P20\_00546 funded by the Junta de Andalucía and the ERDF A way of making Europe,
and in part by Grant PDC2021-121120-C21 funded by MCIN/AEI /10.13039/501100011033 and by the ``European Union \-NextGenerationEU/PRTR".

\begin{appendix}
\gdef\thesection{\Alph{section}}
\makeatletter
\renewcommand\@seccntformat[1]{Appendix \csname the#1\endcsname.\hspace{0.5em}}
\makeatother

\section{Proof of the recursive feasibility of the RMPC controller} \label{app:RMPC:feasibility:proof}

\noindent The proof of Theorem \ref{theo:RMPC:feasibility} makes use of the following property, taken from claims (ii) and (v) of \cite[Theorem 2.1]{KolmanovskyMPE98}.

\begin{property} \label{prop:sets}
Let $\setA, \setB, \setC \subset \R^n$. Then,
\blista
    \item $( \setA \ominus \setB ) \oplus \setB \subseteq \setA$,
    \item $\setA \ominus ( \setB \oplus \setC ) \subseteq ( \setA \ominus \setB) \ominus \setC$.
\elista
\end{property}

\begin{proof}[Proof of Theorem \ref{theo:RMPC:feasibility}] 

Let the optimal solution of $\mathcal{P}_N(x(k))$ be $\bar \vu^* = \left( \bu^*(0; x(k)), \bu^*(1; x(k)), \dots, \bu^*(N-1; x(k)) \right)$ and let $\bar \vx^* = \left( \bx^*(0; x(k)), \bx^*(1; x(k)), \dots, \bx^*(N; x(k)) \right)$ be the corresponding optimal values of the predicted states. 
   
Let us denote a candidate solution of $\mathcal{P}_N(x(k+1))$, where $x(k+1) = A x(k) + B \bu^*(0; x(k)) + w(k)$ is the successor state, by $\bar \vu = \left( \bu(0; x(k+1)), \dots, \bu(N-1; x(k+1)) \right)$ and by $\bar \vx = \left( \bx(0; x(k+1)), \dots, \bx(N; x(k+1)) \right)$ its corresponding predicted states. Additionally, let
\begin{equation} \label{eq:def:fi}
    \delta(i) \doteq \bx(i;x(k+1))-\bx^*(i+1;x(k)), \; i \in \Z_0^{N-1}.
\end{equation}

We construct the candidate solution as follows. First,
\begin{subequations} \label{eq:proof:feasibility:candidate}
\begin{equation} \label{eq:proof:feasibility:candidate:x0}
    \bx(0; x(k+1)) = x(k+1).
\end{equation}
Then, take
\begin{equation} \label{eq:proof:feasibility:candidate:ui}
    \bu(i; x(k+1)) = \bu^*(i+1; x(k)) + K \delta(i), \; i \in \Z_0^{N-2},
\end{equation}
\begin{equation} \label{eq:proof:feasibility:candidate:xi}
    \bx(i; x(k+1)) = A \bx(i-1; x(k+1)) + B \bu(i-1; x(k+1)),
\end{equation}
for $i \in \Z_1^N$, where, finally,
\begin{equation} \label{eq:proof:feasibility:candidate:uN}
    \bu(N-1; x(k+1)) = K_t \bx(N-1; x(k+1)).
\end{equation}
\end{subequations}
Figure \ref{fig:feasibility:trajectory} illustrates the resulting trajectory $\bx(i; x(k+1))$.

In the following, we show that the candidate solution \eqref{eq:proof:feasibility:candidate} is a feasible solution of $\mathcal{P}_N(x(k+1))$ for any $w(k) \in \setW$. Indeed, \eqref{eq:RMPC:initial} and \eqref{eq:RMPC:model} are trivially satisfied from \eqref{eq:proof:feasibility:candidate:x0} and \eqref{eq:proof:feasibility:candidate:xi}.
Next, from the definition \eqref{eq:def:fi} we have that
\begin{align} \label{eq:proof:feasibility:fi_next}
    \delta(i+1) &= \bar x(i+1; x(k+1)) - \bar x^*(i + 2; x(k)) \nonumber\\
           &= A \bar x(i; x(k+1)) + B \bar u(i; x(k+1)) \nonumber\\
           &\quad - A \bar x^*(i+1; x(k)) - B \bar u^*(i+1; x(k)) \nonumber \\
           &= (A + B K) \left( \bar x(i; x(k+1)) - \bar x^*(i+1; x(k)) \right) \nonumber \\
           &= (A + B K) \delta(i) = A_K \delta(i)
\end{align}
for $i \in \Z_0^{N-2}$. Then, noting that
\begin{equation*}
    \delta(0) \numeq{\eqref{eq:proof:feasibility:candidate:x0}} x(k+1) - (A x(k) + B u(k)) = w(k),
\end{equation*}
and recursively applying \eqref{eq:proof:feasibility:fi_next}, we obtain
\begin{equation} \label{eq:proof:feasibility:x_diff}
    \bar x(i; x(k+1)) - \bar x^* (i+1; x(k)) = A_K^i w(k), \; i \in \Z_0^{N-1},
\end{equation}
which by definition of $\setL(\cdot)$ implies that
\begin{equation} \label{eq:proof:feasibility:relation:x}
    \bar x(i; x(k+1)) \in \bar x^* (i+1; x(k)) \oplus \mathcal{L}(i+1), \; i \in \Z_0^{N-1}.
\end{equation}
Therefore, the satisfaction of \eqref{eq:RMPC:const:x} follows from
\begin{align} \label{eq:proof:feasibility:satisfaction:x:N_1}
    \bx(i;x(k+1)) & \in \bx^*(i+1;x(k)) \oplus \setL(i+1) \nonumber \\
                  & \numeq[\subset]{(*)} ( \setX \ominus \setH(i+1) ) \oplus \setL(i+1) \nonumber \\
                  & \numeq[\subseteq]{(**)} \big( ( \setX \ominus \setH(i) ) \ominus \setL(i+1) \big) \oplus \setL(i+1) \nonumber \\
                  & \moveEq{-18} \numeq[\subseteq]{Prop. \ref{prop:sets}(i)} \setX \ominus \setH(i), \; i \in \Z_0^{N-1},
\end{align}
where step (*) follows from the fact that $\bx^*(i;x(k))$ satisfies \eqref{eq:RMPC:const:x} for $i \in \Z_0^{N-1}$ and
\begin{align*}
    &\bx^*(N;x(k)) \numeq[\in]{\eqref{eq:RMPC:terminal}} \Omega_{K_t} \ominus \setL(N) \\
    &\moveEq{-6}\numeq[\subseteq]{\eqref{eq:ass:RMPC:omega:admissible}} (\setX \ominus \setH(N)) \ominus \setL(N) \subseteq \setX \ominus \setH(N),
\end{align*}
and step (**) follows from
\begin{align*}
    &\setX \ominus \setH(i + 1) = \setX \ominus \left( \bigoplus_{j = 0}^i A_K^j \setW \right) \\
    &= \setX \ominus \left( \setH(i) \oplus A_K^i \setW \right) \numeq[\subseteq]{Prop. \ref{prop:sets}(ii)} (\setX \ominus \setH(i) ) \ominus A_K^i \setW.
\end{align*}
From \eqref{eq:proof:feasibility:candidate:ui}, taking into account \eqref{eq:def:fi} and \eqref{eq:proof:feasibility:x_diff}, we have
\begin{equation} \label{eq:proof:feasibility:u_diff}
    \moveEq{-10}\bu(i; x(k+1)) {=} \bu^*(i+1; x(k)) {+} K A_K^i w(k), \; i \in \Z_0^{N-2},
\end{equation}
which, following the same procedure used before, leads to
\begin{align} \label{eq:proof:feasibility:satisfaction:u:N_2}
\bu(i;x(k+1)) & \in \bu^*(i+1;x(k)) \oplus K\setL(i+1) \nonumber\\
              & \subset ( \setU \ominus K\setH(i+1) ) \oplus K\setL(i+1) \nonumber\\
              & \subseteq \big( ( \setU \ominus K\setH(i) ) \ominus K\setL(i{+}1) \big) \oplus K\setL(i{+}1) \nonumber \\
    & \subseteq \setU \ominus K\setH(i), \; i \in \Z_0^{N-2}.
\end{align}

Next, taking $i = N-1$ in \eqref{eq:proof:feasibility:relation:x} leads to
\begin{align} \label{eq:proof:feasibility:x:N_1:in:Omega}
    \bx(N-1;x(k+1)) &\in \bx^*(N;x(k)) \oplus \setL(N) \nonumber \\
                    &\numeq[\subset]{\eqref{eq:RMPC:terminal}} ( \Omega_{K_t} \ominus \setL(N) ) \oplus \setL(N) \nonumber \\
                    &\moveEq{-20}\numeq[\subseteq]{Prop. \ref{prop:sets}(i)} \Omega_{K_t}.
\end{align}
Therefore, taking into account the definition of the Pontryagin difference, we have that
\begin{align*}
    \bx(N; x(k{+}1)) &= A \bx(N{-}1; x(k{+}1)) {+} B \bu(N{-}1; x(k{+}1)) \\
                     &\moveEq{-5}\numeq{\eqref{eq:proof:feasibility:candidate:uN}} (A + B K_t) \bx(N-1; x(k+1)) \\
                     &\moveEq{-1}\numeq[\subseteq]{\eqref{eq:proof:feasibility:x:N_1:in:Omega}} (A + B K_t) \Omega_{K_t} \\
                     &\moveEq{2}\numeq[\subseteq]{\eqref{eq:ass:RMPC:omega:invariant}} \Omega_{K_t} \ominus \setL(N),
\end{align*}
which shows the satisfaction of \eqref{eq:RMPC:terminal}.

Finally, since $\bx(N-1; x(k+1)) \in \Omega_{K_t}$ \eqref{eq:proof:feasibility:x:N_1:in:Omega}, and taking into account Assumption \ref{ass:RMPC}.(iv), we have that
\begin{align*}
    \bu(N-1; x(k+1)) &\moveEq{-6}\numeq{\eqref{eq:proof:feasibility:candidate:uN}} K_t \bx(N-1; x(k+1)) \\
                     &\numeq[\in]{\eqref{eq:ass:RMPC:omega:admissible}} \setU \ominus K \setH(N-1)
\end{align*}
which, alongside \eqref{eq:proof:feasibility:satisfaction:u:N_2}, proves the satisfaction of \eqref{eq:RMPC:const:u}. \qedhere

\end{proof}

\section{Proof of the input-to-state stability of the RMPC controller} \label{app:RMPC:ISS:proof}

\noindent The proof of Theorem \ref{theo:RMPC:ISS} makes use of the following proposition.

\begin{proposition} \label{prop:quadratic:bound:dif}
    Let $\setC$ be a compact set, $a, b, c \in \R^n$ satisfy $a = b + M c$ with $M \in \R^{n \times n}$ and $b \in \setC$. Then, for any given $S \succ 0 \in \R^{n \times n}$ there exists a $\setK_\infty$ function $\rho(\cdot)$ such that 
\begin{equation*}
    \| a \|_S^2 - \| b \|_S^2 \leq \rho \left( \| c \| \right).
\end{equation*}
\end{proposition}

\begin{proof}
Denote $\tau = \max_{b \in \setC} \| M\T S b \|$. We have that,
\begin{align*}
    \| a \|_S^2 - \| b \|_S^2 &= \| b + M c \|_S^2 - \| b \|_S^2 \\
                              &= 2 b\T S M c + c\T M\T S M c \\
                              &\numeq[\leq]{(*)} 2 \| M\T S b \| \|c \| + \lambda_{\text{max}}( M\T S M\T ) \| c \|^2 \\
                              &\leq 2 \tau \| c \| + \lambda_{\text{max}}( M\T S M\T ) \| c \|^2,
\end{align*}
where $(*)$ is due to the Cauchy-Schwarz inequality.
Thus, the claim of the property holds with
\begin{equation*}
    \rho(\| c \| ) = 2 \tau \| c \| + \lambda_{\text{max}}( M\T S M\T ) \| c \|^2. \qedhere
\end{equation*}
\end{proof}

\begin{proof}[Proof of Theorem \ref{theo:RMPC:ISS}] 

    In the following, we will prove that the optimal cost function of $\mathcal{P}_N(x(k))$, $V^*_N(x(k))$, is an ISS Lyapunov function of the closed-loop system for any ${w(k) \in \setW}$.

    Let the optimal solution of problem $\mathcal{P}_N(x(k))$ be given by $\bar \vu^* = \left( \bu^*(0), \dots, \bu^*(N{-}1) \right)$ and $\bar \vx^* = \left( \bx^*(0), \dots, \bx^*(N) \right)$ be the corresponding optimal values of the predicted states. Additionally, consider the successor state 
\begin{equation*}
    x(k+1) = A x(k) + B \bu^*(0; x(k)) + w(k),
\end{equation*}
and let us denote by $\bar \vu$ and $\bar \vx$ the trajectories given by \eqref{eq:proof:feasibility:candidate} for this successor state, which by Theorem \ref{theo:RMPC:feasibility} are a feasible solution of $\mathcal{P}_N(x(k+1))$.
    
We note that, as shown in the above definitions of $\bar \vu^*$ and $\bar \vx^*$, we are dropping the notation $\bu^*(i; x(k))$ for $\bu^*(i)$ to improve readability. In fact, we will drop the $(k)$ notation entirely. Instead, the variables referring to the successor state will be indicated with a superscript $+$, as in $x^+ \equiv x(k+1)$.
    
To prove the claim we follow \cite[Theorem 3]{LimonLNCIS09}, which states that the existence of $\alpha_1$, $\alpha_2$, $\alpha_3$ $\setK_\infty$-functions and a $\setK$-function $\alpha_4$ such that
\begin{subequations} \label{eq:ISS:Lyapunov:conditions}
\begin{align}
    \moveEq{-8}&V^*_N(x) \geq \alpha_1(\|x\|),\, \forall x\in \setX, \label{eq:ISS:Lyapunov:conditions:lower} \\
    \moveEq{-8}&V^*_N(x) \leq \alpha_2(\|x\|),\, \forall x\in \Omega_{K_t}, \label{eq:ISS:Lyapunov:conditions:upper} \\
    \moveEq{-8}&V_N^*(x^+) {-} V_N^*(x) {\leq} {-}\alpha_3(\|x\|) {+} \alpha_4(\|w\|), \forall x{\in}\setX, w{\in}\setW, \label{eq:ISS:Lyapunov:conditions:diff}
\end{align}
\end{subequations}
proves that $V^*_N(x)$ is an ISS Lyapunov function of the closed loop system.

First, we show that the lower and upper bounds \eqref{eq:ISS:Lyapunov:conditions:lower} and \eqref{eq:ISS:Lyapunov:conditions:upper} can be obtained following standard procedures, see \cite{MayneAUT00}. Indeed, the lower bound \eqref{eq:ISS:Lyapunov:conditions:lower} can be obtained by taking into account that $Q \succ 0$ as follows,
\begin{equation*}
    V^*_N(x) \geq \|\bar x^*(0) \|^2_Q = x\T Q x \geq \lambda_{\text{min}}(Q) \|x\|^2, \forall x \in \setX.
\end{equation*}
Whereas the upper bound \eqref{eq:ISS:Lyapunov:conditions:upper} can be obtained in the terminal region as follows,
\begin{equation*}
    V^*_N(x)\leq x^TPx \leq \lambda_{\text{max}}(P)\|x\|^2, \forall x \in \Omega_{K_t},
\end{equation*}
where the left-hand side inequality is a well known result in the field of MPC for all states $x \in \Omega_{K_t}$.

To prove the satisfaction of \eqref{eq:ISS:Lyapunov:conditions:diff}, let us first note the following inequality,
\begin{align} \label{eq:proof:stability:bound:N_1_usefull}
    & \| \bx^+(N-1)\|^2_Q + \| \bu^+(N-1) \|^2_R + \| \bx^+(N) \|^2_P \nonumber \\
    &\numeq{\eqref{eq:proof:feasibility:candidate:xi}} \| \bx^+(N-1)\|^2_Q + \| \bu^+(N-1) \|^2_R \nonumber \\
    &\quad\;\;+ \| A \bx^+(N-1) + B \bu^+(N-1) \|^2_P \nonumber \\
    &\numeq{\eqref{eq:proof:feasibility:candidate:uN}} \| \bx^+(N-1) \|^2_{Q + K_t\T R K_t + (A + B K_t)\T P (A + B K_t)} \nonumber \\
    &\;\numeq[\leq]{\eqref{eq:ass:RMPC:P}} \| \bx^+(N-1) \|^2_P.
\end{align}

Then, we have that
\begin{align} \label{eq:proof:stability:cost:x_plus}
    V_N(x^+) &= \sum\limits_{i=0}^{N-2} \big( \| \bx^+(i) \|^2_Q + \| \bu^+(i) \|^2_R \big) {+} \| \bx^+(N) \|^2_P\nonumber \\
             &\quad+ \| \bx^+(N{-}1) \|_Q^2 + \| \bu^+(N{-}1) \|^2_R  \nonumber \\
    &\numeq[\leq]{\eqref{eq:proof:stability:bound:N_1_usefull}} \sum\limits_{i=0}^{N-2} \big( \| \bx^+(i) \|^2_Q + \| \bu^+(i) \|^2_R \big) {+} \| \bx^+(N{-}1) \|^2_P.
\end{align}
Additionally, eliminating $\| \bu^*(0) \|^2_R$ from $V_N^*(x)$ leads to
\begin{align} \label{eq:proof:stability:cost:x}
    V_N^*(x) \geq &\sum\limits_{i = 0}^{N-2} \big( \| \bx^*(i+1) \|^2_Q + \| \bu^*(i+1) \|^2_R \big) \nonumber \\
                  &+ \| \bx^*(0) \|^2_Q + \| \bx^*(N) \|^2_P.
\end{align}
Therefore, from \eqref{eq:proof:stability:cost:x_plus} and \eqref{eq:proof:stability:cost:x}, and noting that $\bx^*(0) = x$, we have that
\begin{align} \label{eq:proof:stability:cost:original}
    \moveEq{-0}V_N(x^+) {-} V_N^*(x)\, {\leq} & \sum\limits_{i=0}^{N-2} \big( \| \bx^+(i) \|^2_Q - \| \bx^*(i{+}1) \|^2_Q \big) \nonumber \\
                                          &+ \sum\limits_{i = 0}^{N-2} \big( \| \bu^+(i) \|^2_R - \| \bu^*(i{+}1) \|^2_R \big) \nonumber \\
                                          &- \| x \|^2_Q + \| \bx^+(N{-}1) \|^2_P - \| \bx^*(N) \|^2_P.
\end{align}

Next, from \eqref{eq:proof:feasibility:x_diff} we know that
\begin{equation*}
    \bx^+(i) = \bx^*(i+1) + A_K^i w, \, i \in \Z_0^{N-1},
\end{equation*}
for any $w \in \setW$. Then, since $\bx^*(i+1)$ belongs to a compact set, we have from Proposition \ref{prop:quadratic:bound:dif} that,
\begin{subequations} \label{eq:proof:stability:bound:x_i}
\begin{align} 
    &\|\bx^+(i)\|^2_Q-\|\bx^*(i+1)\|^2_Q \leq \alpha_i ( \| w \| ), \, i \in \Z_0^{N-2}, \\
    &\|\bx^+(N-1)\|^2_P-\|\bx^*(N)\|^2_P \leq \alpha_{N-1} ( \| w \| ),
\end{align}
\end{subequations}
for some $\setK_\infty$ functions $\alpha_i(\cdot)$, $i \in \Z_0^{N-1}$.
Similarly, from \eqref{eq:proof:feasibility:u_diff}, and taking into account that $\bu^*(i+1)$ belongs to a compact set, we have that
\begin{equation} \label{eq:proof:stability:bound:u_i}
    \| \bu^+(i) \|^2_R - \| \bu^*(i+1) \|^2_R \leq \gamma_i( \| w \| ), \, i \in Z_0^{N-2},
\end{equation}
for some $\setK_\infty$ functions $\gamma_i(\cdot)$, $i \in \Z_0^{N-2}$.

Therefore, using \eqref{eq:proof:stability:bound:x_i} and \eqref{eq:proof:stability:bound:u_i} in \eqref{eq:proof:stability:cost:original} leads to
\begin{equation*}
    V_N(x^+) - V_N^*(x) \leq - \| x \|^2_Q + \sigma( \| w \| ),
\end{equation*}
where $\sigma(\| w \|)$ is given by
\begin{equation*}
    \sigma(\| w \|) = \sum\limits_{i = 0}^{N-1} \alpha_i(\| w \|) + \sum\limits_{i = 0}^{N-2} \gamma_i(\|w\|),
\end{equation*}
which is, by construction, a $\setK_\infty$-function.
Finally, by optimality we have $V_N^*(x^+)\leq V_N(x^+)$. Thus,
\begin{equation*}
    V_N^*(x^+) - V_N^*(x) \leq - \| x \|^2_Q + \sigma( \| w \| ). \qedhere
\end{equation*}

\end{proof}

\onecolumn
\section{Extended case study results for the double reactor and separator plant} \label{app:extended}

\begin{figure*}[h!]
    \centering
    \begin{subfigure}[ht]{0.32\textwidth}
        \includegraphics[width=\linewidth]{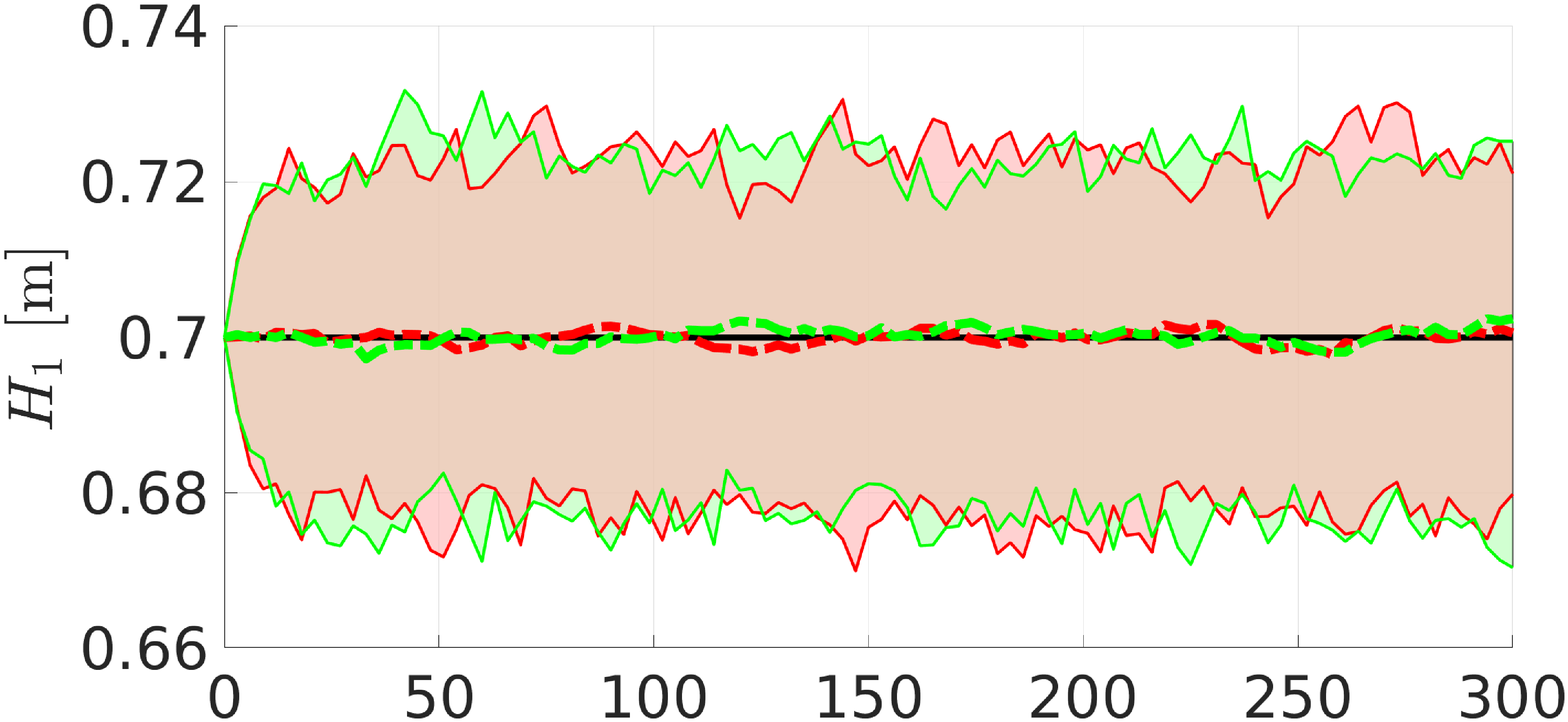}
    \end{subfigure}%
    \;
    \begin{subfigure}[ht]{0.32\textwidth}
        \includegraphics[width=\linewidth]{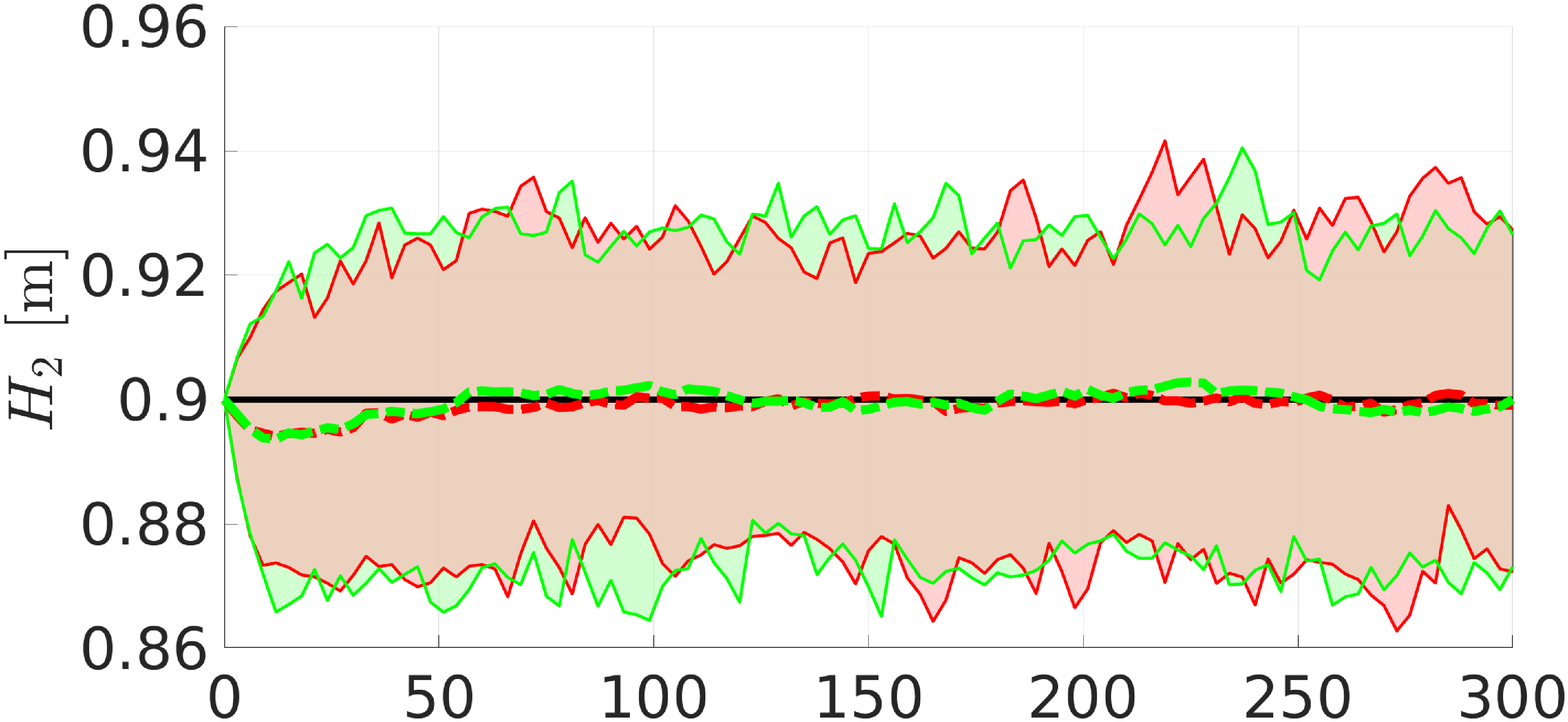}
    \end{subfigure}%
    \;
    \begin{subfigure}[ht]{0.32\textwidth}
        \includegraphics[width=\linewidth]{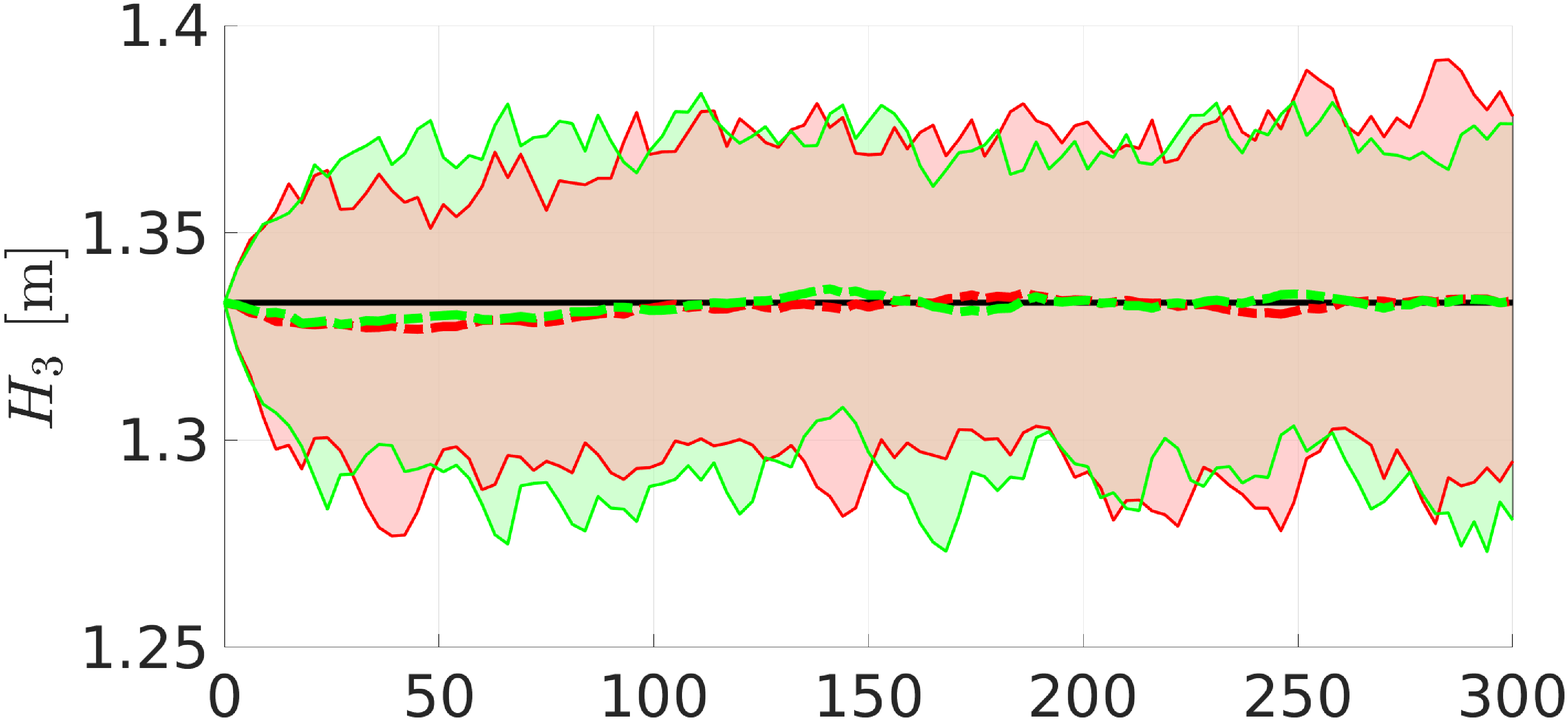}
    \end{subfigure}%
    \vspace{1em}

    \begin{subfigure}[ht]{0.32\textwidth}
        \includegraphics[width=\linewidth]{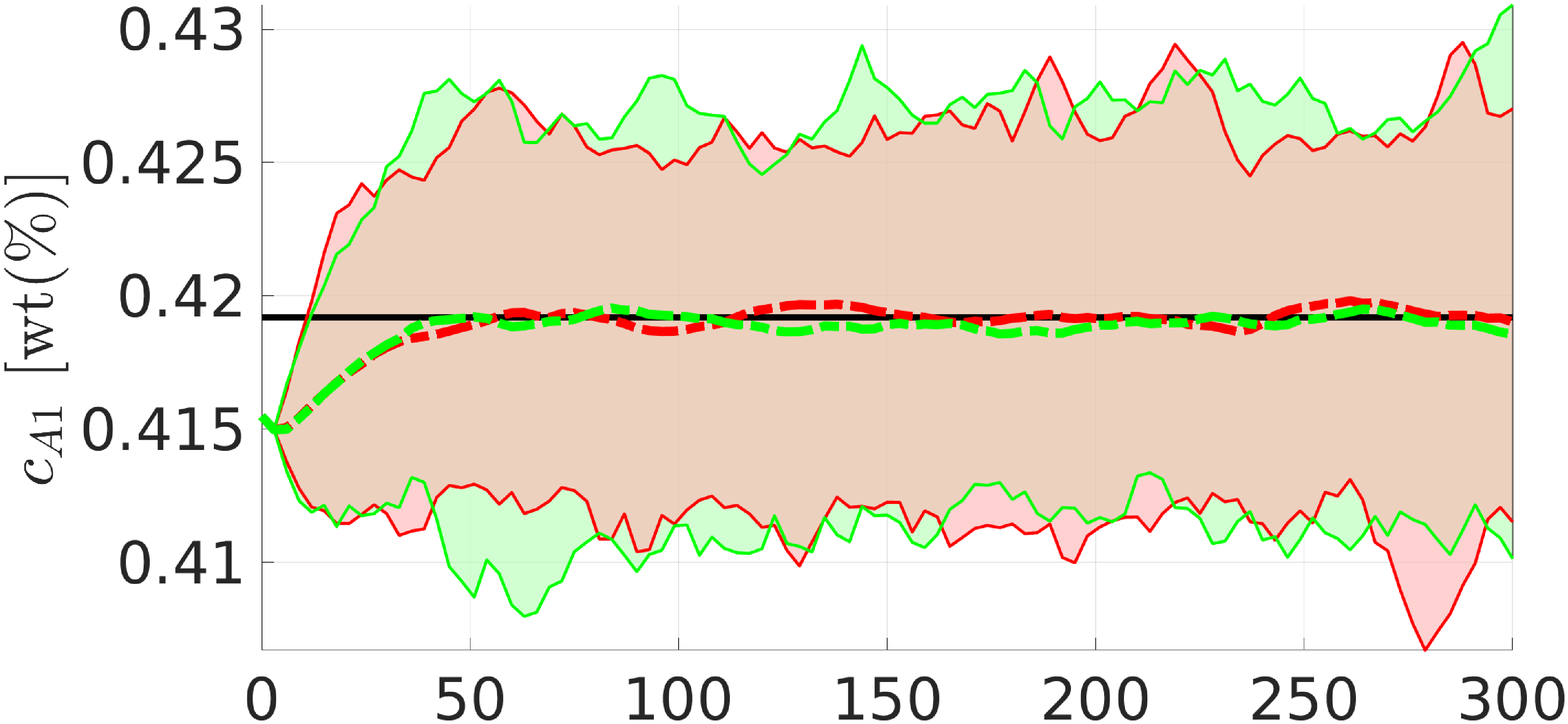}
    \end{subfigure}%
    \;
    \begin{subfigure}[ht]{0.32\textwidth}
        \includegraphics[width=\linewidth]{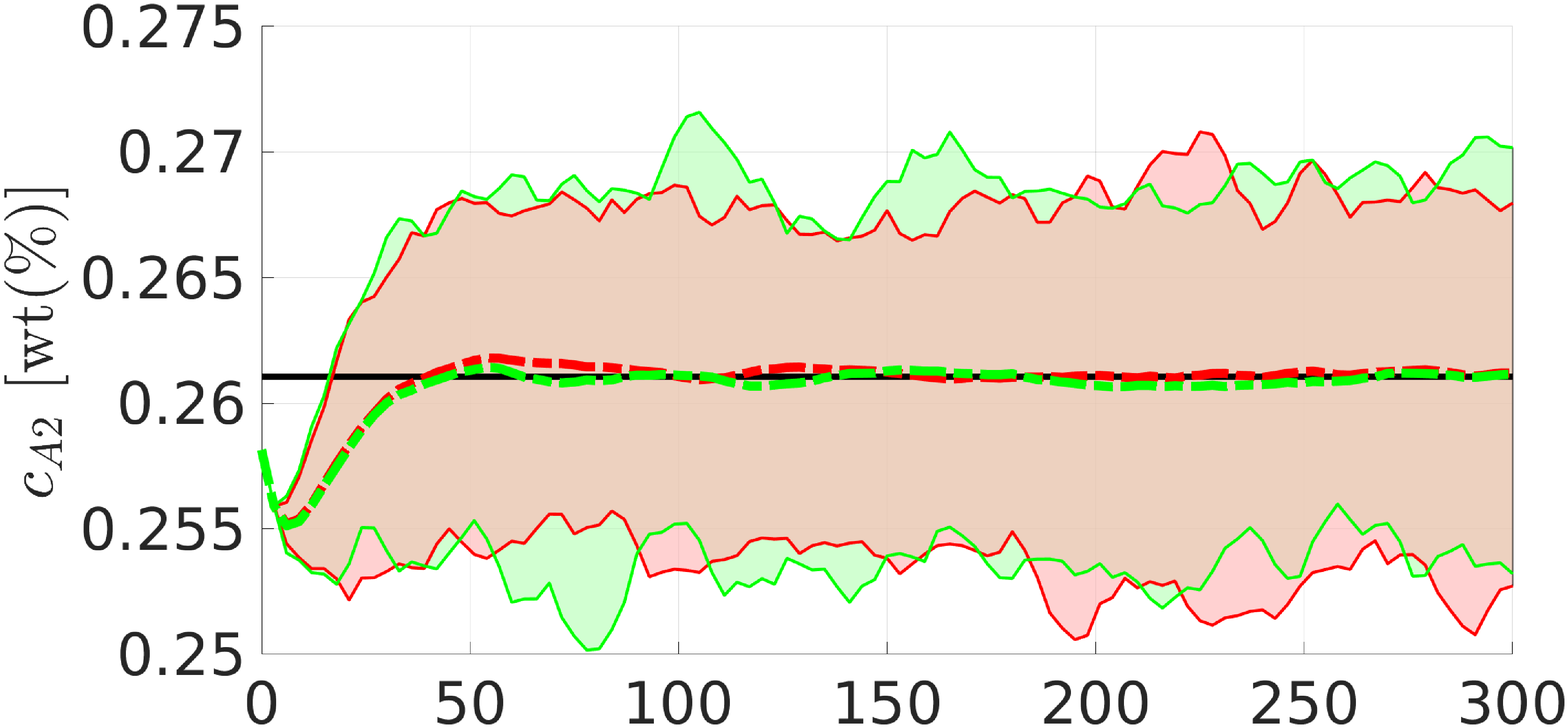}
    \end{subfigure}%
    \;
    \begin{subfigure}[ht]{0.32\textwidth}
        \includegraphics[width=\linewidth]{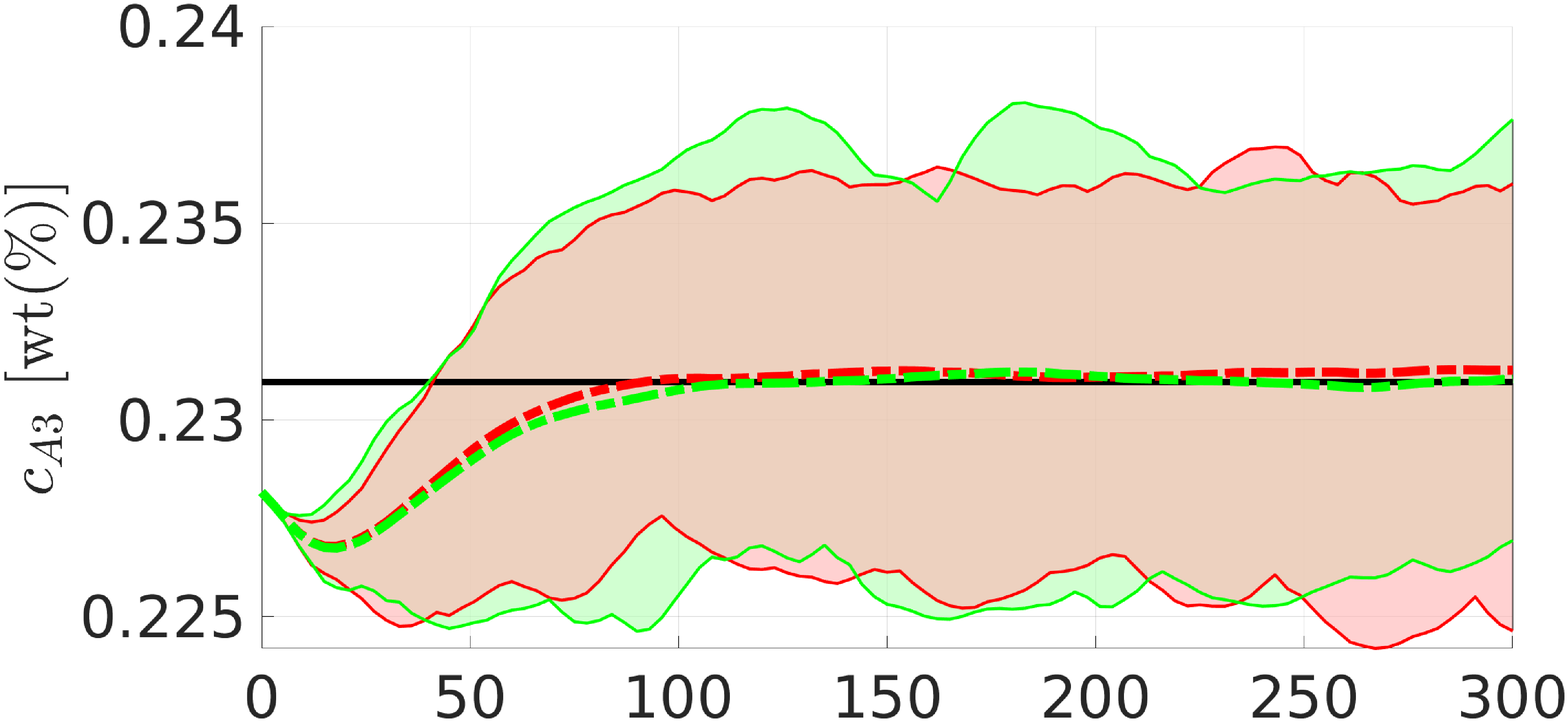}
    \end{subfigure}%
    \vspace{1em}

    \begin{subfigure}[ht]{0.32\textwidth}
        \includegraphics[width=\linewidth]{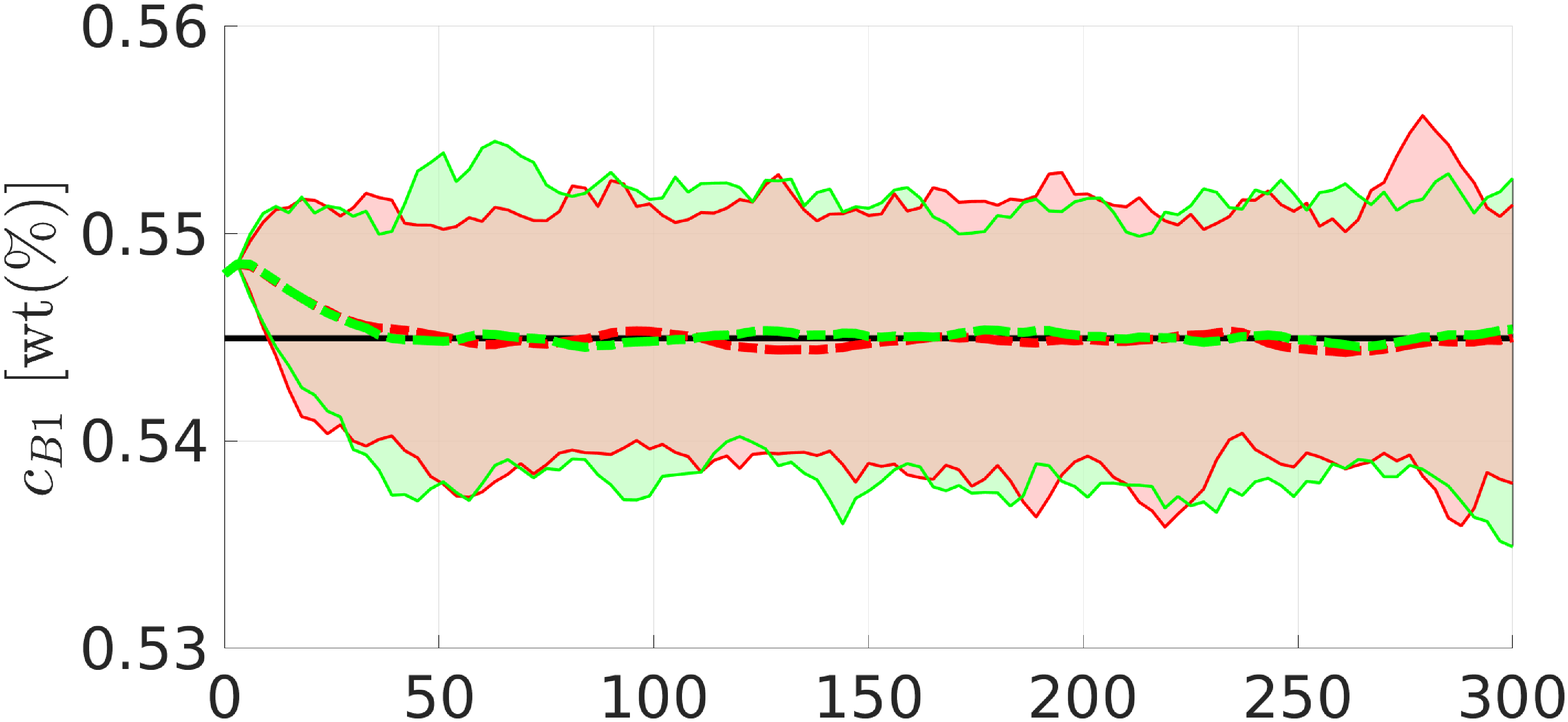}
    \end{subfigure}%
    \;
    \begin{subfigure}[ht]{0.32\textwidth}
        \includegraphics[width=\linewidth]{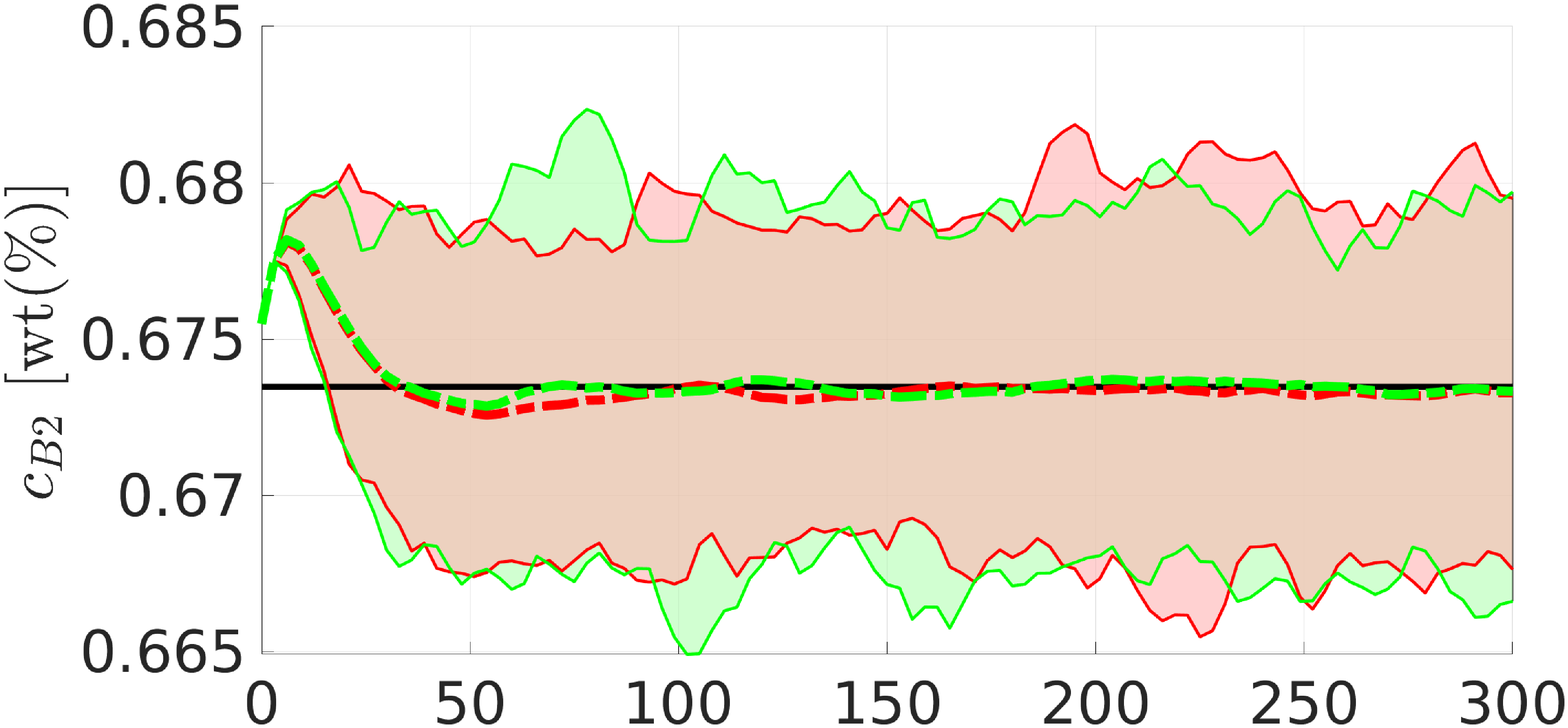}
    \end{subfigure}%
    \;
    \begin{subfigure}[ht]{0.32\textwidth}
        \includegraphics[width=\linewidth]{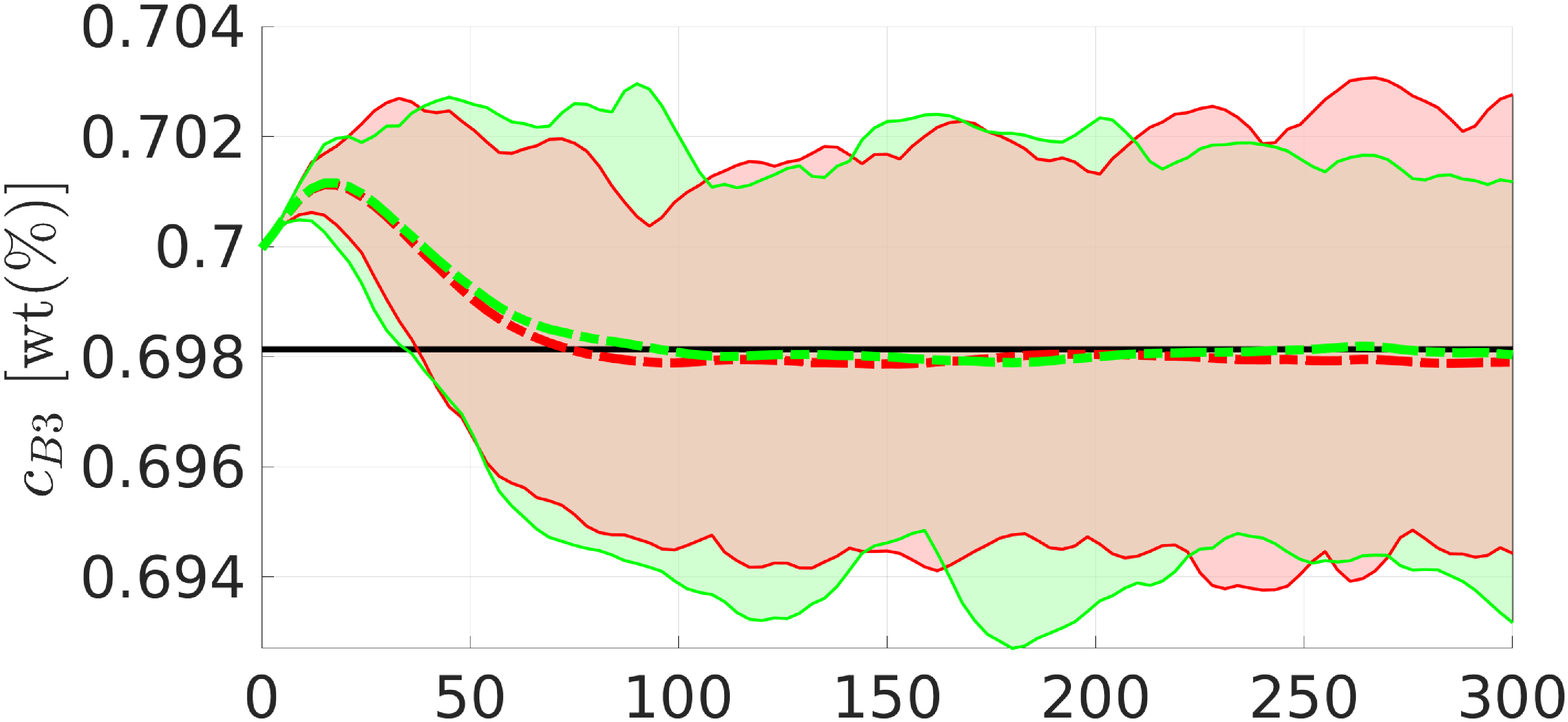}
    \end{subfigure}%
    \vspace{1em}

    \begin{subfigure}[ht]{0.32\textwidth}
        \includegraphics[width=\linewidth]{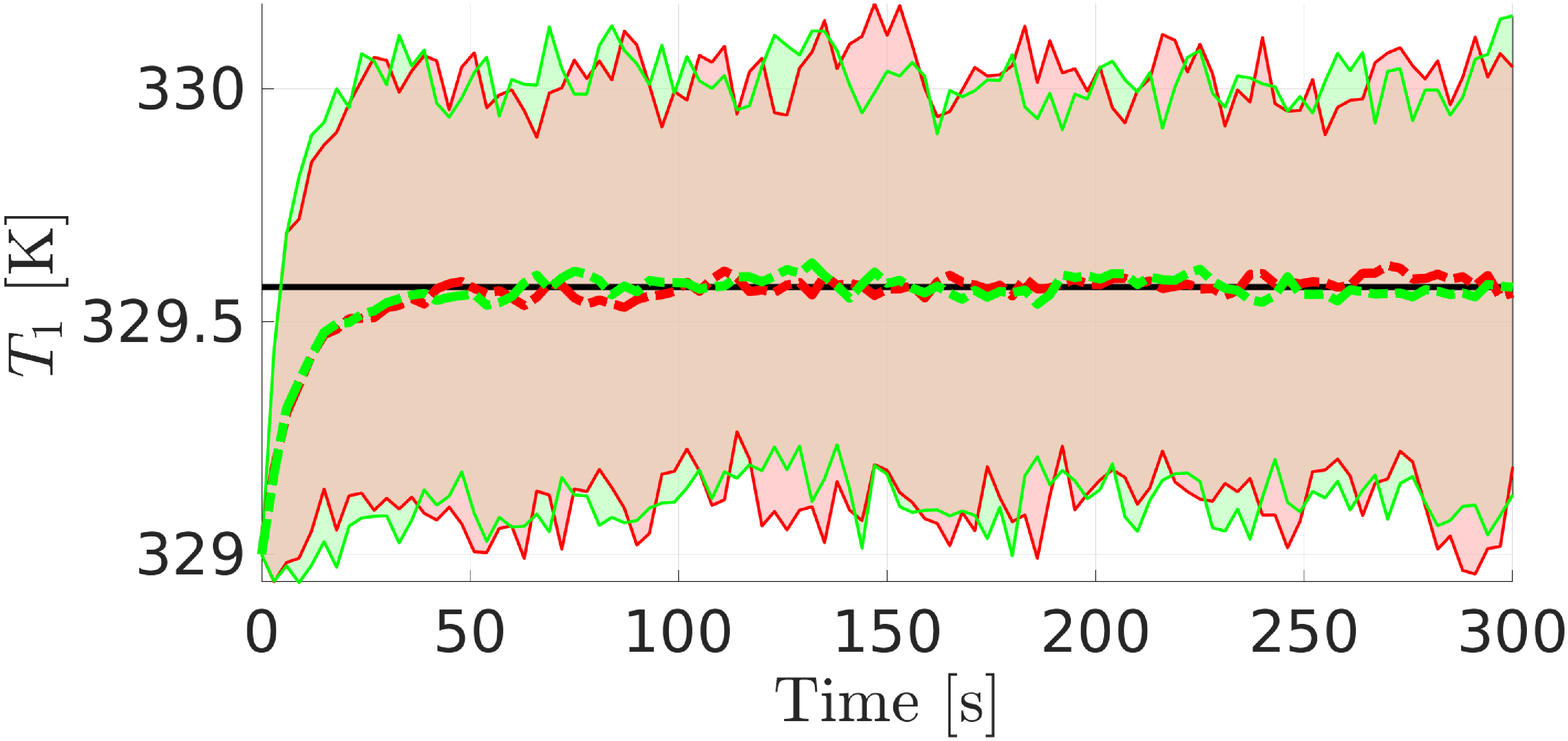}
    \end{subfigure}%
    \;
    \begin{subfigure}[ht]{0.32\textwidth}
        \includegraphics[width=\linewidth]{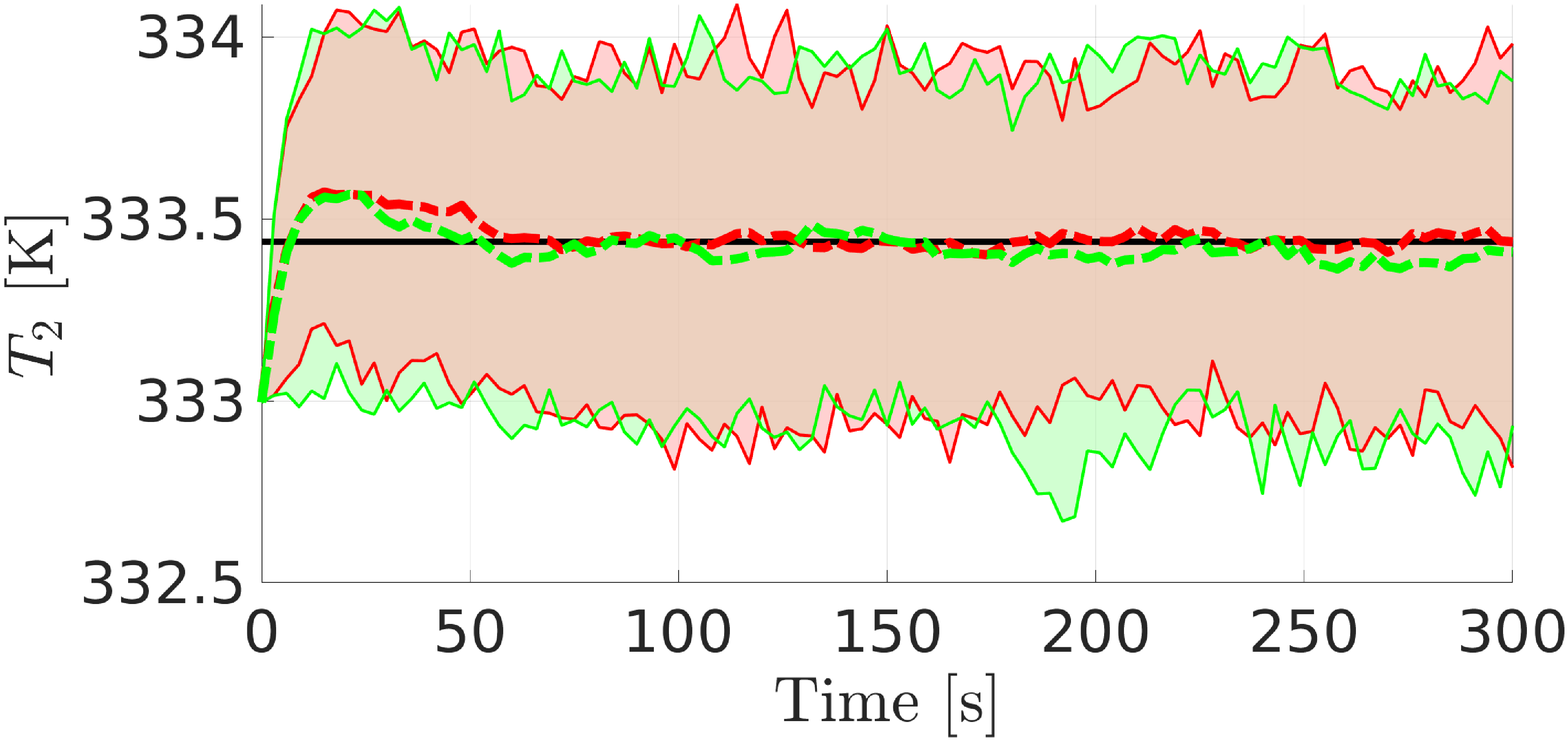}
    \end{subfigure}%
    \;
    \begin{subfigure}[ht]{0.32\textwidth}
        \includegraphics[width=\linewidth]{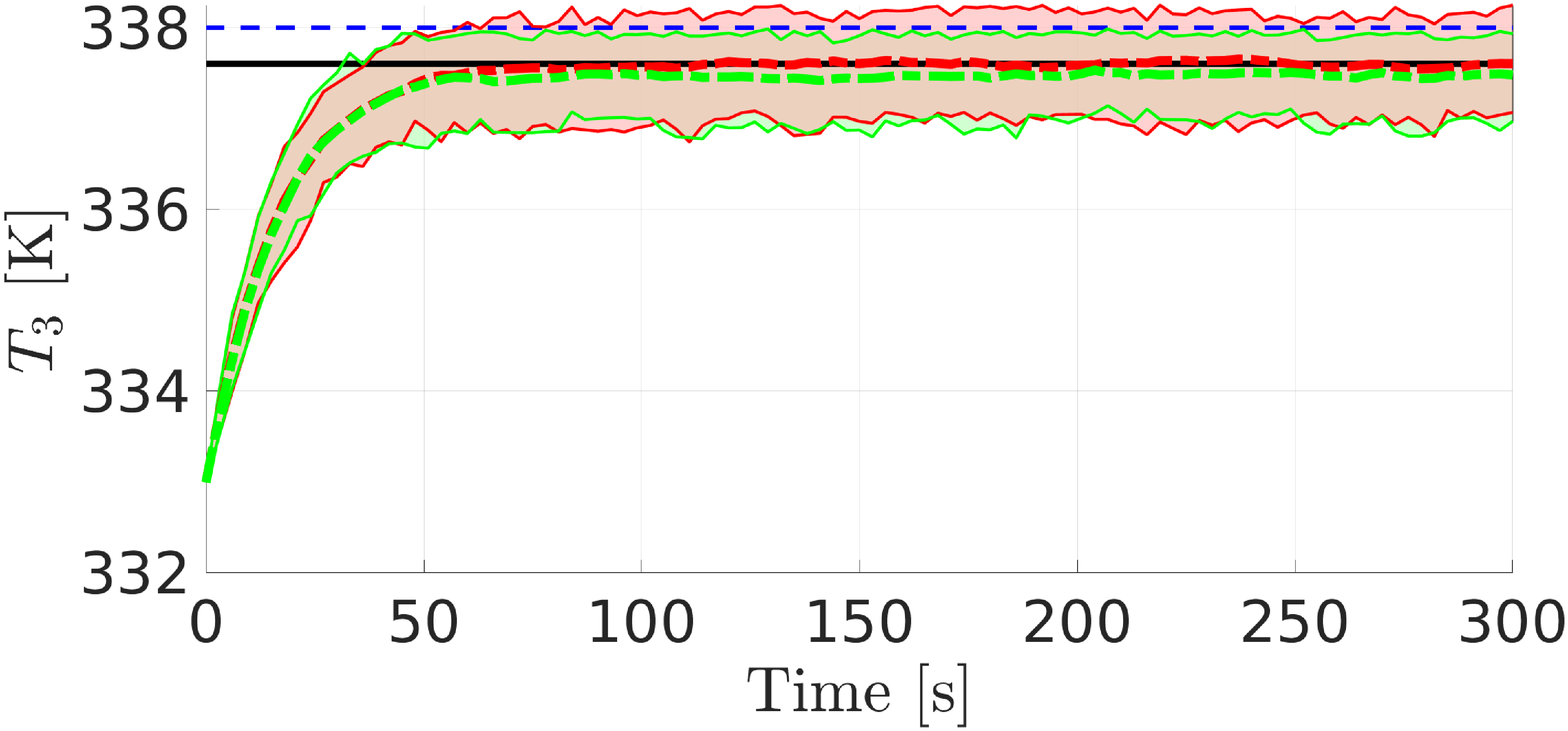}
    \end{subfigure}%

    \caption{States during the test shown in Section \ref{sec:case:study:plant} for the double reactor and separator plant.
Reference in solid back line and state constraint in dashed blue line. Otherwise, the lines and shaded areas represent the same as in Figure \ref{fig:reactors:tests}, but representing the results for the proposed RMPC in green and the results for the nominal MPC in red.
We only represent the upper constraint of $T_3$ because all other constraints are inactive during all the simulations.}
    \label{fig:extended:state}
    \vspace{2em}
    \begin{subfigure}[ht]{0.32\textwidth}
        \includegraphics[width=\linewidth]{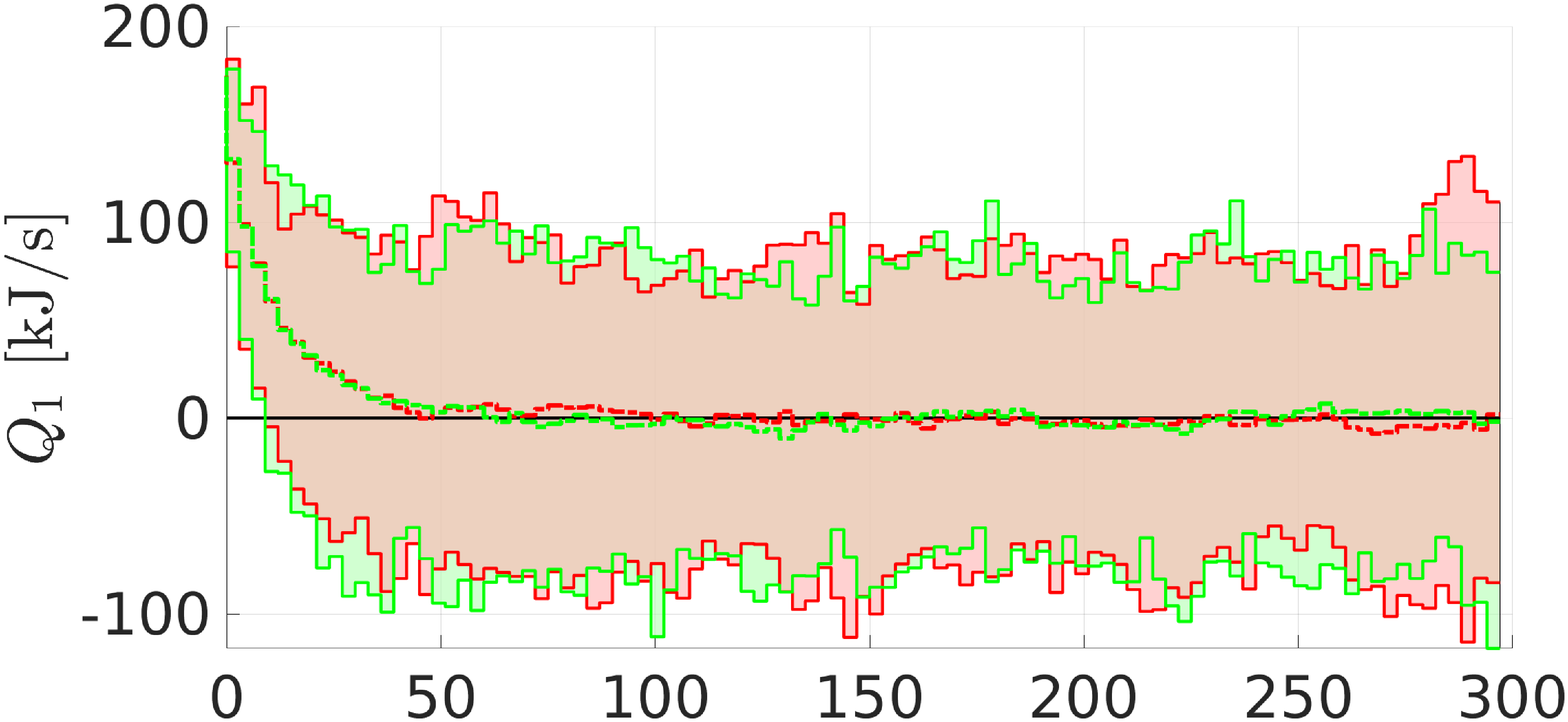}
    \end{subfigure}%
    \;
    \begin{subfigure}[ht]{0.32\textwidth}
        \includegraphics[width=\linewidth]{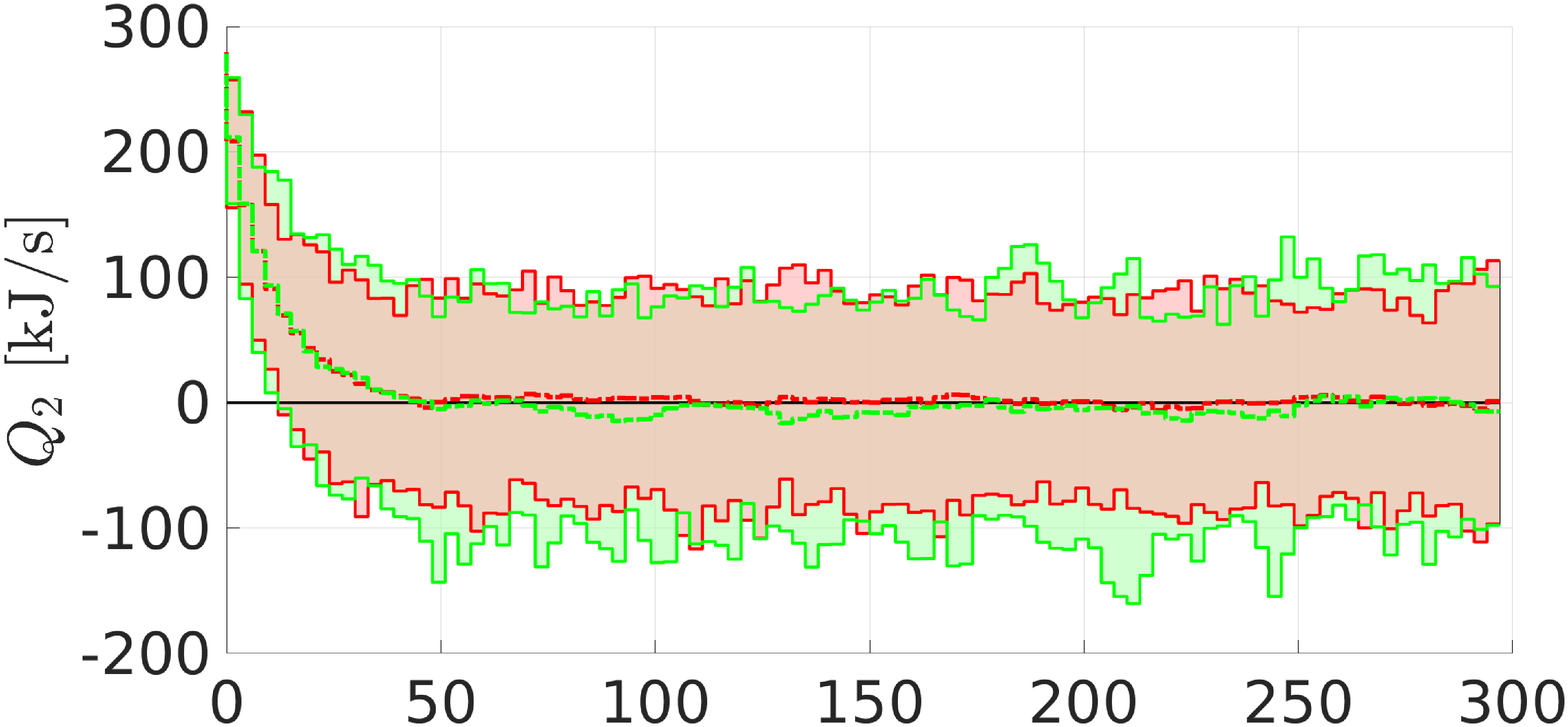}
    \end{subfigure}%
    \;
    \begin{subfigure}[ht]{0.32\textwidth}
        \includegraphics[width=\linewidth]{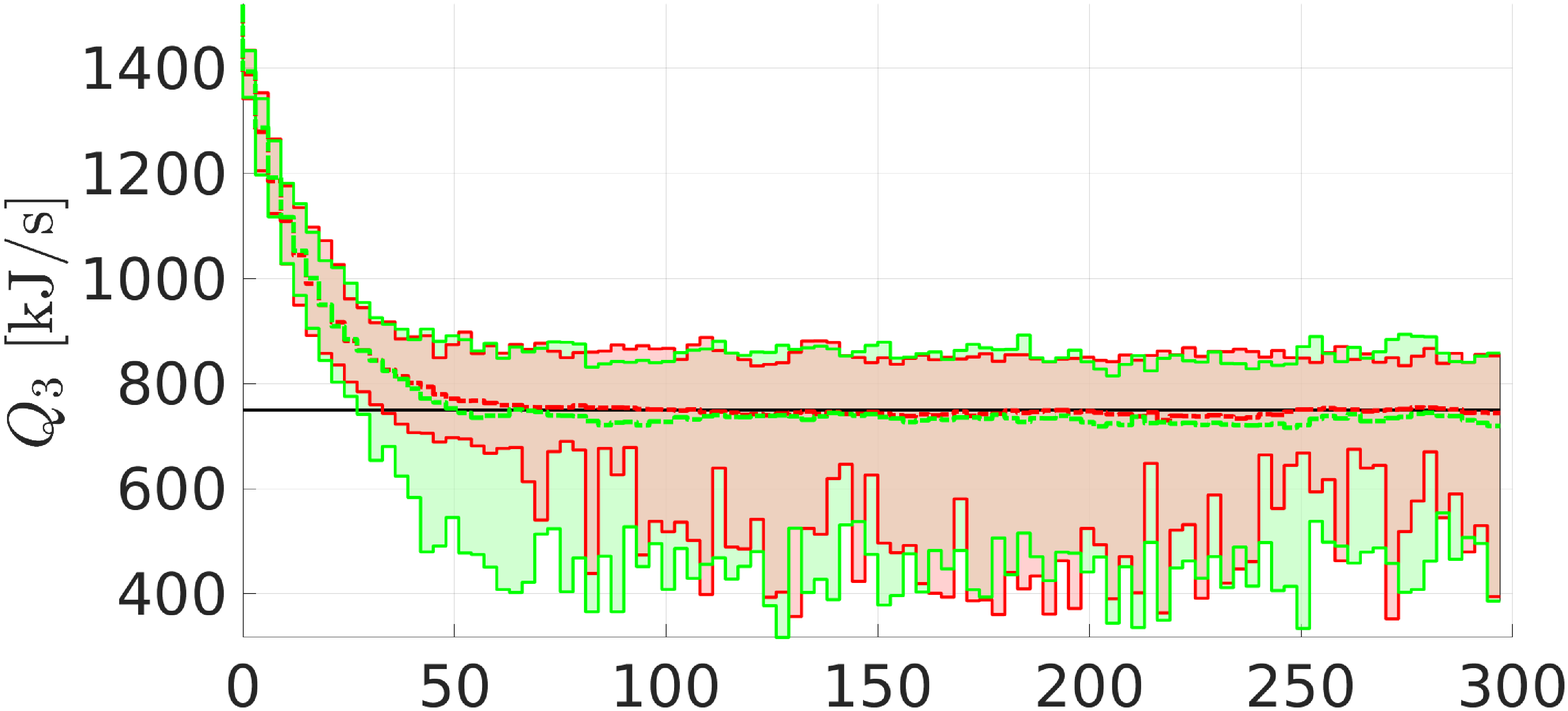}
    \end{subfigure}%
    \vspace{1em}

    \begin{subfigure}[ht]{0.32\textwidth}
        \includegraphics[width=\linewidth]{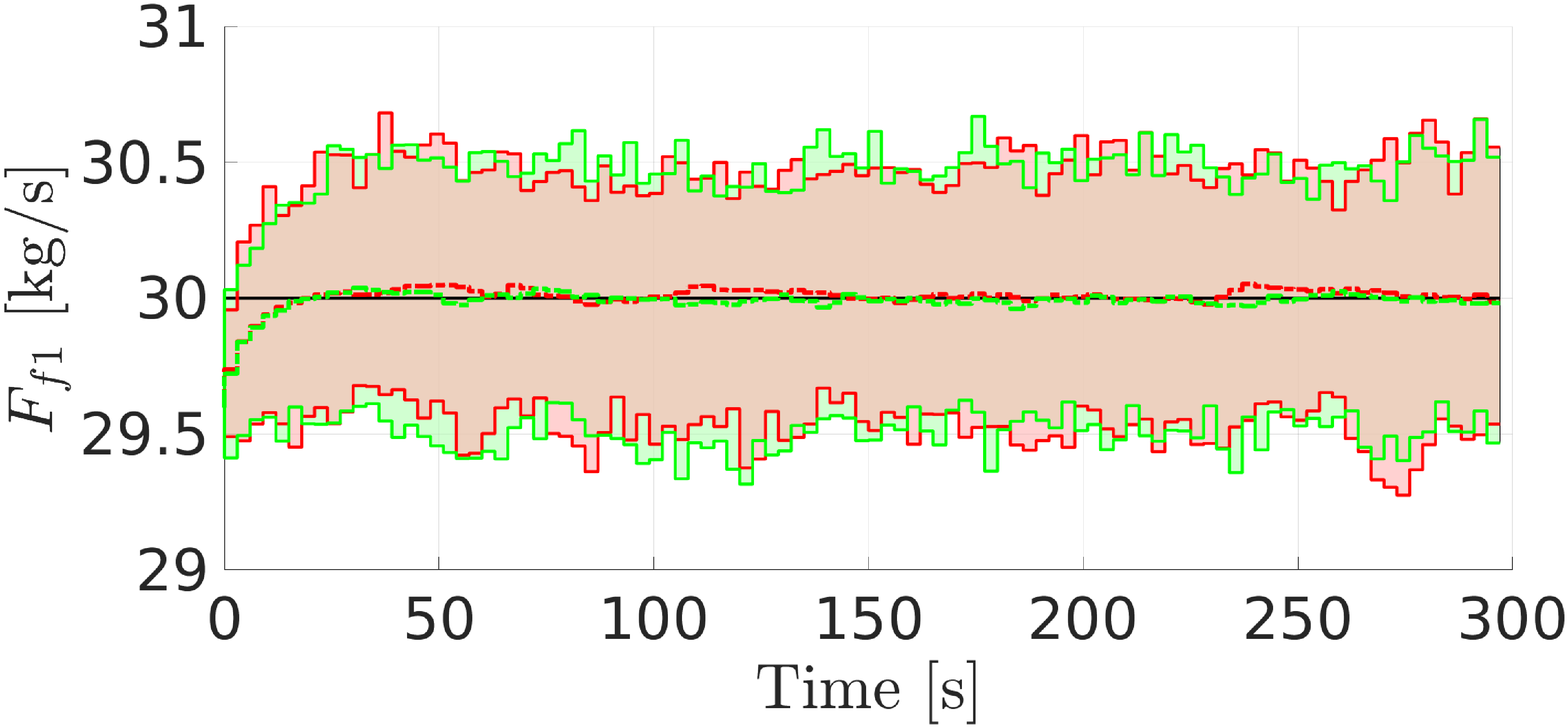}
    \end{subfigure}%
    \;
    \begin{subfigure}[ht]{0.32\textwidth}
        \includegraphics[width=\linewidth]{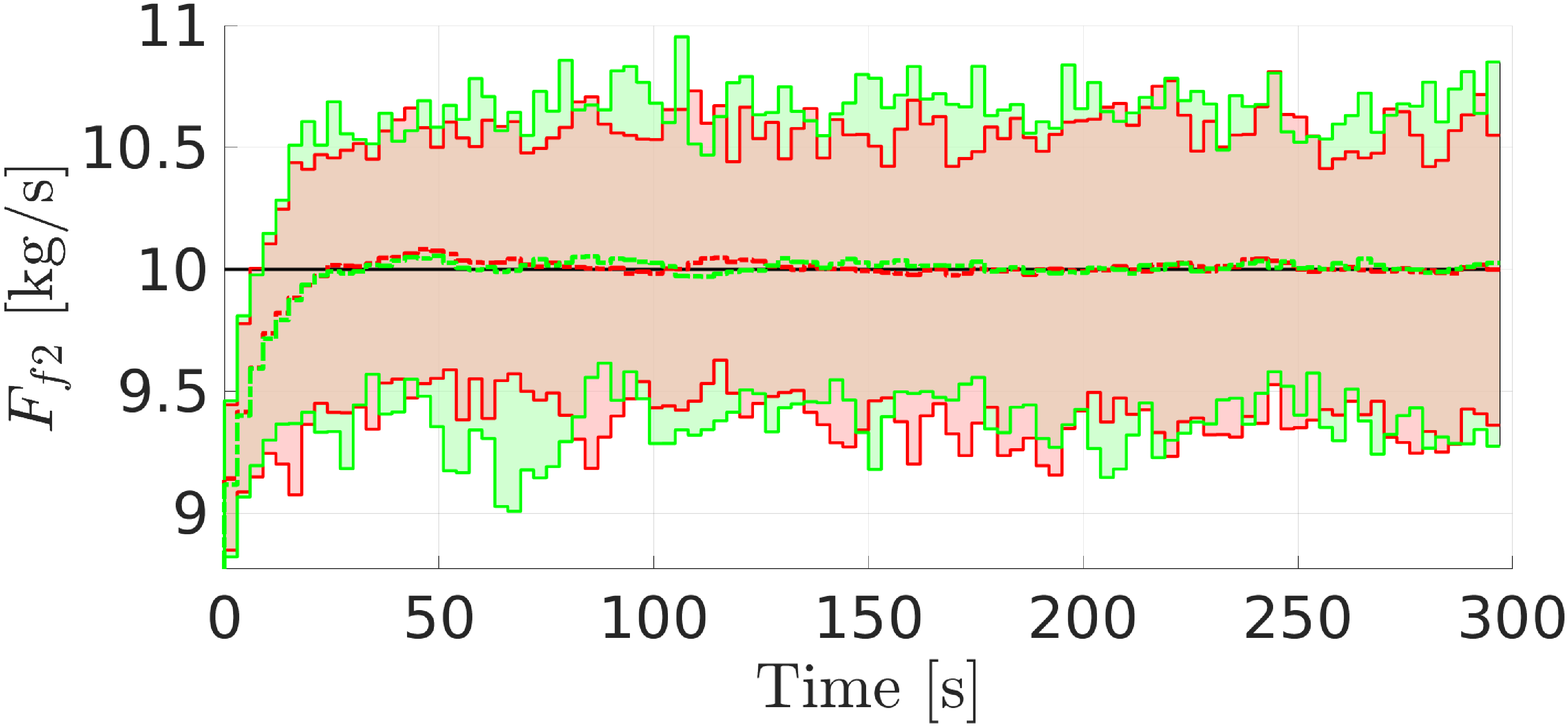}
    \end{subfigure}%
    \;
    \begin{subfigure}[ht]{0.32\textwidth}
        \includegraphics[width=\linewidth]{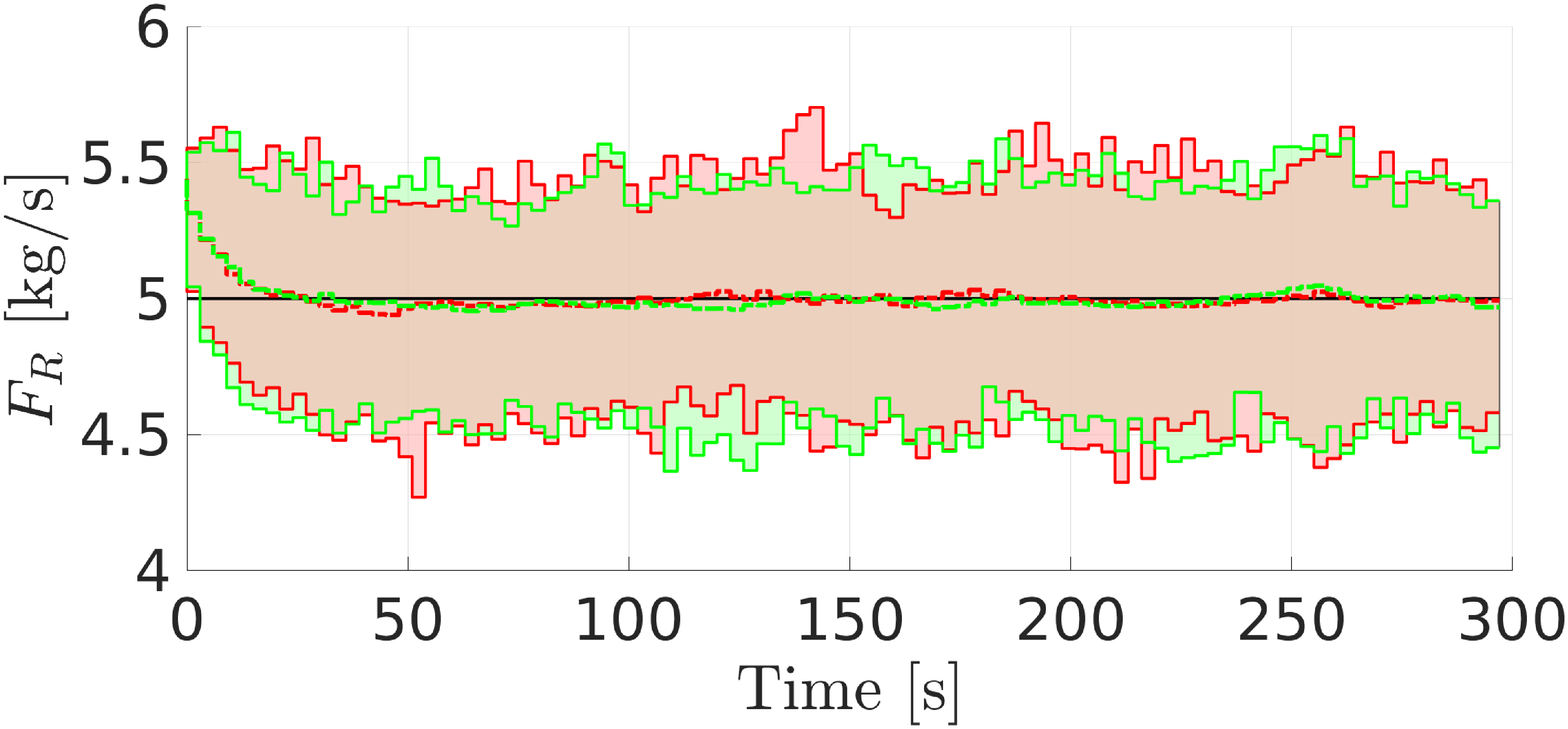}
    \end{subfigure}%

    \caption{Control inputs during the test shown in Section \ref{sec:case:study:plant} for the double reactor and separator plant.
Reference in solid black line.
Otherwise the lines and shaded areas represent the same as in Figures \ref{fig:reactors:tests} and \ref{fig:extended:state}, where green is used to represent the results using the proposed RMPC and red using the nominal MPC.
We do not represent the constraints because they are inactive during all the simulations.}
\end{figure*}
\vspace*{-2em}

\twocolumn

\end{appendix}

\bibliographystyle{elsarticle-num}
\bibliography{IEEEabrv,tesis}

\end{document}